\documentclass[5p]{elsarticle}
\PassOptionsToPackage{hyphens}{url}
\usepackage{xcolor, siunitx, booktabs}
\usepackage[colorlinks]{hyperref}
\usepackage[tableposition=top]{caption}
\usepackage{float}
\usepackage{amsfonts} 
\usepackage{breqn}
\usepackage{algorithm2e}
\usepackage{algorithmic}
\usepackage{tabularx}
\usepackage{supertabular}
\usepackage{ragged2e}
\DeclareMathOperator{\NER}{NER}
\DeclareMathOperator{\DBSCAN}{DBSCAN}
\title{A large dataset of software mentions in the biomedical literature}
\author[1]{Ana-Maria Istrate}
\ead{aistrate@chanzuckerberg.com}
\author[1]{Donghui Li}
\ead{dli@chanzuckerberg.com}
\author[1]{Dario Taraborelli}
\ead{dario@chanzuckerberg.com}
\author[1]{Michaela Torkar}
\ead{michaela.torkar@contractor.chanzuckerberg.com}
\ead{dario@chanzuckerberg.com}
\author[1,2]{Boris Veytsman}
\ead{bveytsman@chanzuckerberg.com}
\author[1]{Ivana Williams}
\ead{iwilliams@chanzuckerberg.com}

\affiliation[1]{organization={Chan Zuckerberg Initiative},
  addressline={801 Jefferson Street},
  city={Redwood City},
  state={CA},
  postcode={94063},
  country={USA}}

\affiliation[2]{organization={George Mason University},
addressline={Fairfax},
state={VA},
postcode={22030},
country={USA}}

\begin{document}
\begin{abstract}
  We describe the CZ Software Mentions dataset, a new dataset of
  software mentions in biomedical papers.  Plain-text software
  mentions are extracted with a trained SciBERT model from several
  sources: the NIH PubMed Central collection and from papers provided
  by various publishers to the Chan Zuckerberg Initiative. The dataset
  provides sources, context and metadata, and, for a number of
  mentions, the disambiguated software entities and links. We extract
  1.12 million unique string software mentions from 2.4 million papers
  in the NIH PMC-OA \textsl{Commercial} subset, 481k unique mentions
  from the NIH PMC-OA \emph{Non-Commercial} subset (both gathered in
  October~2021) and 934k unique mentions from 3~million papers in the
  \emph{Publishers' collection}.  There is variation in how software
  is mentioned in papers and extracted by the NER algorithm. We
  propose a clustering-based disambiguation algorithm to map
  plain-text software mentions into distinct software entities and
  apply it on the NIH PubMed Central \textsl{Commercial}
  collection. Through this methodology, we disambiguate 1.12 million
  unique strings extracted by the NER model into \num{97600} unique
  software entities, covering 78\% of all software-paper links. We
  link \num{185000} of the mentions to a repository, covering about
  55\% of all software-paper links. We describe in detail the process
  of building the datasets, disambiguating and linking the software
  mentions, as well as opportunities and challenges that come with a
  dataset of this size. We make all data and code publicly available
  as a new resource to help assess the impact of software (in
  particular scientific open source projects) on science.

\end{abstract}

\maketitle
\thispagestyle{plain}
\tableofcontents
\listoffigures
\listoftables
\listofalgorithms

\section{Introduction}
\label{sec:intro}

The common adage says that the work of the scientist is only as good
as their tools. Since the last century software has become a key tool
in a scientist’s toolbox and in recent years some of the most
important breakthroughs in science---from the solution of 50-year-old
protein folding problem~\cite{Jumper2021} to the first-ever direct
image of a black hole's event horizon~\cite{Sagittarius22}---have been
made possible through advanced computational methods applied to large
swaths of data. Yet, identifying and crediting the computational tools
that enabled these discoveries, and rewarding their creators, remains
a challenge. While there are established norms in science for giving
formal citation and credit to the authors of scholarly papers going
back centuries, software---as many types of non-traditional research
outputs~\cite{Priem2010, Bucknell2016, Alperin2022}---is often
neglected or treated as a second-class type of output and eminently
hard to cite. Not being able to measure the impact of critical
software tools that enable scientific progress makes it hard for their
authors and maintainers to pursue scientific careers and to obtain
funding for their work~\cite{Howison2011, Howison2015,
  SinghChawla2016, Howison2016, Knowles2021}. Furthermore, it makes it more
difficult for other scientists to reproduce results in scientific
papers, and creates barriers for funders who need to objectively
evaluate the impact of their support~\cite{Howison2016, Mesirov2010,
  Joppa2013, Nowogrodzki2019, Strasser2022}.   

While these problems have been well recognized \cite{Katz2014,
  Howison2016}, and a number of recommendations for citing software
have been published \cite{Smith2016, Katz2021}, in scientific papers
software is often only credited with informal mentions and phrases like \emph{for analysis
  ImageJ software was used.}  Until citation practices for software
are improved, an analysis of the text of the papers remains necessary
in order to get insight into software usage.

The extraction of software mentions from the text of scientific papers
has attracted the attention of researchers for quite some time (see
the review \cite{Krueger2020} and the recent publications
\cite{Lopez2021, Orduna-Malea2021}).  There are several datasets
available, including \textsl{SoftwareKG}, a knowledge graph that
contains information about software mentions from more than
\num{51000} scientific articles from the social sciences
\cite{Schindler2020}; \textsl{SoMeSci}, a curated collection of
\num{3756}~software mentions in \num{1367}~PubMed Central articles
\cite{Schindler2021}; \textsl{Softcite}, a dataset of manual
annotations of \num{4971}~academic PDFs in biomedicine and economics
\cite{Du2021}; \cite{Lopez2021dataset}, a dataset of
\num{318138}~software mentions based on CORD-19 dataset; and
\cite{Wade2021}, another dataset of \num{77449}~software mentions
based on CORD-19 dataset.  These datasets are based on limited samples
of scholarly articles.  A paper based on a similar
approach~\cite{Schindler2022} was published while this preprint was in
preparation.  Similarly to our work, it is based on Pubmed Central
Open Access dataset. The methodology for software extraction is
similar and based on SciBERT~\cite{beltagy2019scibert}. The paper uses
a hierarchical multi-task labeling model to extract additional tags
besides software and the version, such as software type, mention type
and additional information.  The training corpora for our datasets are
different: we trained our model on SoftCite~\cite{Du2021}, while
\citeauthor{Schindler2022} trained the model on the SoMeSci
dataset~\cite{Schindler2021}.  Our disambiguation methodologies are
overall similar and built on clustering algorithms, with differences
in the type of features used.  There is also a difference in the
number of clusters obtained.  Our linking methodology is quite
different. The authors predict links based on training on the SoMeSci
dataset, whereas we predict links by querying a number of repositories
on names of clusters obtained through disambiguation. The authors
extract \num{301825757} triples describing \num{11.8}M software
mentions and construct a knowledge graph based on the results.  The
sizes of the resulting mention datasets are similar: ours has
\num{19.3}M mentions (Table~\ref{tab:oapmcstats}).  The formats of the
final datasets are also different: for example, we provide context for
each mention in the dataset.  We plan to conduct a more detailed
comparison of our results and methods.

In this work we describe a series of larger datasets of software
mentions based on (1)~the full PubMed Central collection
\cite{PMCHelp} downloaded in October~2021 containing more than
3.8~million papers (Table~\ref{tab:oapmcstats}), and (2)~a collection
of papers provided by various publishers to the Chan Zuckerberg
Initiative (Table~\ref{tab:publishers}).  We use a SciBERT-based model
trained on the SoftCite dataset~\cite{Du2021} to extract software
mentions from these corpora. There can be large variability in how a
software entity is mentioned in text or extracted by the NER model, so
we propose a clustering-based technique to disambiguate the plain-text
mentions into distinct software entities
(Section~\ref{sec:disamb}). We also describe a methodology to link the
mentions to PyPI, CRAN, Bioconductor, SciCrunch or GitHub
(Section~\ref{sec:linking}).

We apply the disambiguation and linking methodologies to the PMC-OA
\textsl{comm} subset, from which the NER model extracts 1.12 million
unique string software mentions appeared in 2.4 million papers. We are
able to disambiguate \num{320000} of these mentions into \num{97600}
unique software entities, covering about 78\% of all links in the
dataset. For all the other strings extracted by the NER model, we
cannot confidently map them to a software entity through our
methodology. We also link about \num{185000} mentions to a repository,
covering about 55\% of all software-paper links.

We curate a proportion of the final datasets by engaging a domain
expert team. Through this process, we find that the precision of the
NER model `in the wild'', on the top 10000 mentions (by frequency) in
the PMC-OA \textsl{comm} subset, or Precision@10k is 69.66\% (Precision score of the model on
the SoftCite dataset is 90\%). We describe in detail our methodology
and make the final datasets available to the community. We hope this
work serves as a resource for assessing the impact of scientific
software.

\section{Materials and Methods}
\label{sec:materials}

\subsection{Full text collection}
\label{sec:collection}

We used two separate full text collections.  The first dataset is
based on the PubMed Central collection \cite{PMCHelp} (\emph{PMC OA)}
downloaded on October 2021.  The PMC OA collection has two subsets:
\textsl{comm} subset licensed for both commercial and non-commercial
use, and \emph{non\_comm} subset licensed for non-commercial use only,
including data mining.  We kept these data separate to ensure that our
data set can be utilized for both commercial and non-commercial uses
(see the discussion of non commercial licensing in
\cite{Veytsman2021}). The statistics of the data set is shown in
Table~\ref{tab:publishersstats}.  Second, there were full-text
manuscripts of scholarly articles (both Open Access and paywalled
manuscripts) provided to the Chan Zuckerberg Initiative by publishers
under different agreements (\emph{Publishers' collection}). This
corpus also includes preprints from \textsl{bioRxiv}.  The provenance
and time span of this collection varies (see
Table~\ref{tab:publishers}).  As seen from Tables~\ref{tab:publishers}
and~\ref{tab:oapmcstats}, the datasets have significant overlap.

The Publishers' collection was stored in LXML format PMC OA papers were
downloaded in NXML format.  For parsing LXML we used \textsl{lxml}
\cite{behnel2005lxml}; for parsing NXML we used a modified
\textsl{pubmed\_parser} software \cite{Achakulvisut2020}.  Our
modifications concerned table caption extraction (not implemented in
the original) and speeding up parsing.  They, along with the other
modules, are available at the GitHub site accompanying this
publication.  The extracted text was fed into the SciBERT-based NER model
(Section~\ref{sec:extraction}) and further processed
(Section~\ref{sec:disamb}).

\begin{table}
  \centering
  \caption{PMC OA statistics}
  \label{tab:oapmcstats}
  \begin{tabularx}{\columnwidth}{X<{\raggedright}*{2}{S[table-format=8]}}
    \toprule
    Parameter & {Commercial set} & {Non-commercial set}\\
    \midrule
    Number of papers & 2433010 & 1442868\\
    Number of papers with at least one mention & 1732603 & 758246 \\
    Number of mentions & 14770209 & 4546607\\
    Number of unique mentions & 1120125 & 481972 \\
    \bottomrule
  \end{tabularx}
\end{table}

\begin{table}
  \centering
  \caption{Publishers collection statistics}
  \label{tab:publishersstats}
  \begin{tabularx}{\columnwidth}{X<{\raggedright}*{1}{S[table-format=8]}}
    \toprule
    Parameter & {Publishers Collection} \\
    \midrule
    Number of papers & 16809266\\
    Number of papers with at least one mention & 2893518 \\
    Number of mentions & 48160836 \\
    Number of unique mentions & 934704 \\
    \bottomrule
  \end{tabularx}
\end{table}

\begin{table*}
  \centering
  \caption{Papers in Publishers' Collection}
  \label{tab:publishers}
  \begin{tabular}{lS[table-format=7]}
    \toprule
    Publisher & \multicolumn{1}{p{10em}}{\centering Number of papers
                with at least one mention}\\ 
    \midrule
American Association of Neurological Surgeons & 7512\\
American College of Physicians & 1001\\
American Institute of Aeronautics and Astronautics & 8670\\
American Institute of Physics & 48565\\
American Physical Society & 9649\\
American Psychiatric Association Publishing & 5171\\
American Society for Clinical Investigation & 8787\\
American Society for Microbiology & 4943\\
American Society of Agricultural and Biological Engineers & 1132\\
American Society of Civil Engineers & 30641\\
American Thoracic Society & 2247\\
Annual Reviews & 8307\\
BioOne & 61822\\
bioRxiv & 33136\\
Cambridge University Press & 205\\
CSIRO Publishing & 5999\\
De Gruyter Open & 4603\\
Edinburgh University Press & 6784\\
eLife Sciences Publications, Ltd & 1652\\
Emerald Publishing Limited & 55367\\
Future Medicine Ltd & 18934\\
Hindawi & 23052\\
Hogrefe Publishing & 8067\\
Impact Journals & 4191\\
INFORMS & 669\\
Institute of Electrical \& Electronics Engineers & 43580\\
IntechOpen & 1\\
International Union of Crystallography & 21\\
IOS Press & 3989\\
MA Healthcare & 8284\\
Mary Ann Liebert, Inc., publishers & 65597\\
MDPI & 14334\\
MIT Press & 2544\\
Public Library of Science & 186313\\
PubMed Central Open Access & 1758247\\
Royal College of Surgeons of England & 1458\\
Royal Society of Chemistry & 73586\\
SAGE Publications & 358849\\
SLACK Incorporated & 2801\\
Society of Photo-Optical Instrumentation Engineers & 31\\
Springer Nature & 329910\\
Taylor \& Francis & 607702\\
The Royal Society & 342\\
University of California Press & 457\\
Wolters Kluwer & 32940\\\addlinespace
Total & 3852092\\
    \bottomrule
  \end{tabular}
\end{table*}

\subsection{Software Mentions Extraction}
\label{sec:extraction}
We use a NER model to extract plain-text software mentions from our
corpus. The SciBERT \cite{beltagy2019scibert} model has been
fine-tuned on the SoftCite dataset \cite{Du2021} to recognize mentions
of software and respective version. This model achieves a 10-fold
cross-validation F1 score of 0.92. More details can be found at the
address
\url{https://github.com/chanzuckerberg/software-mention-extraction}.
This model was previously used on the CORD-19 dataset~\cite{Wade2021}.

For a part of the collection: OA PMC corpus released under free
license for both commercial and non-commercial use (\path{comm}
dataset) we performed linking and disambiguation described below. 

\subsection{Software Mentions Linking and Disambiguation}

\begin{figure}
\centering
\includegraphics[width=\linewidth]{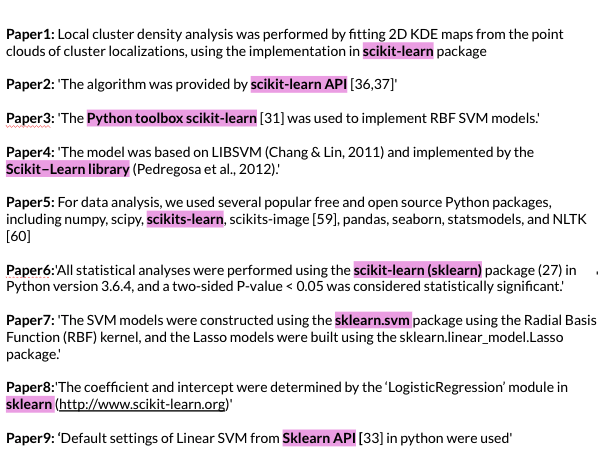}
\caption[Variations of a mention]{Variations of a mentions. All of the
  string variations should be counted as mentions of the same software
  entity. In this case, all of these mentions should be counted as
  mentions of the software entity \emph{scikit-learn}. These examples
  come from running the NER model on the PMC-OA corpus. The
  highlighted sequence is the exact sequence extracted by the NER
  model.}
\label{fig:scikit_learn_example}
\end{figure}

There can be large variability in how a piece of software is being
mentioned in text (Figure~\ref{fig:scikit_learn_example}). For
example, a software can be mentioned through its full name
(e.g. \textit{Statistical Package for Social Sciences}) or its acronym
(e.g. \textit{SPSS}). There can also be multiple name variations
commonly accepted as referring to the same software entity
(e.g. \textit{sklearn} and \textit{scikit-learn}). And there can be
variability in how a software mention is being mentioned by
researchers (eg \textit{Image J} and \textit{ImageJ} or
\textit{GraphPad Prism} and \textit{GraphPad} and
\textit{Prism}). Moreover, there can be typos, either introduced by
the authors (e.g. \textit{scikits-learn}) or by parsing the XML of the
papers, and there can be variability in how the NER algorithm extracts
different software mentions (e.g. partial matches).  If we want to
assess the impact of a software entity, we need to be able to map all
of the different string variations that correspond to a particular
software entity together.

\begin{figure}
\centering
\includegraphics[width=\linewidth]{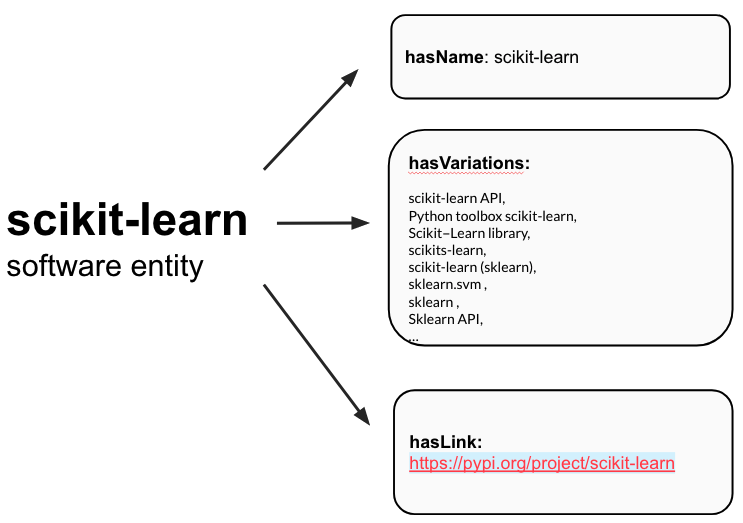}
\caption{Software entity data model}
\label{fig:data_model}
\end{figure}

Our goal is to build a software entity data model and describe a
software entity through its name (eg \emph{scikit-learn},
Figure~\ref{fig:data_model}), the string variations under which this
entity appears mentioned in the literature, as extracted by the NER
model (e.g. \textit{sklearn, Sklearn API, scikit-learn API,
  scikits-learn, Python toolbox scikit-learn, Scikit-Learn Library,
  etc}) and, ideally, a link to a repository or database, such as
\url{https://pypi.org/project/scikit-learn}.

To achieve this, after \emph{extracting} the software mentions from
the PMC-OA corpus using the NER model, we use two additional
methodologies: \emph{disambiguation} and \emph{linking} for the part
free for commercial use (Table~\ref{tab:oapmcstats}). We define
disambiguation as the process of mapping various string variations of
the same software entity together. Linking refers to mapping a
software entity to a URL in a repository or database. Our algorithm
is shown as Algorithm~\ref{alg:disambiguation}. We describe each of
these steps in more detail in the following sections.

\begin{algorithm}
  \caption{Disambiguation and linking}
  \label{alg:disambiguation}
\textbf{Input} Papers $P = P_1, P_2, ... P_n$ \\
\textbf{Output} $\text{software\_entities} \leftarrow \{\}$ \\
\begin{algorithmic} 
\FOR{$P_i\in$P}
   \STATE $V_i \leftarrow \NER(P_i)$ // strings extracted by the NER algorithm from $P_i$
\ENDFOR
\STATE $S \leftarrow V_1 \cup V_2 \cup ... \cup V_n$ // set of all strings extracted
\STATE $M \leftarrow \text{get\_similarity\_matrix}(S)$
\STATE $C_1, C_2, ... C_n \leftarrow \text{get\_connected\_components}(M)$
\FOR{$C_i\in C_1, C_2, ... C_n$}
   \STATE $M_i \leftarrow \text{get\_similarity\_matrix}(C_i)$
   \STATE $D_i \leftarrow 1 - M_i$ //distance matrix
   \FOR{$\text{Cluster}_j{_{D_i}}\in \DBSCAN(D_i)$}
    \STATE $s_{j_{D_i}} \leftarrow \text{highest\_frequency\_string}(\text{Cluster}_j{_{D_i}})$
    \STATE $\text{software\_entities}[s_{j_{D_i}}][\text{\ttfamily\small 'variations'}] \leftarrow \text{Cluster}_j{_{D_i}}$
    \STATE
    $\text{software\_entities}[s_{j_{D_i}}][\text{\ttfamily\small 'link'}] \leftarrow \text{get\_link}(s_{j_{D_i}})$
  \ENDFOR
\ENDFOR
\end{algorithmic}
\end{algorithm}

\subsection{Software Mentions Disambiguation}
\label{sec:disamb}

In the PMC-OA \textsl{comm} corpus, \emph{MATLAB}, the
\emph{R package limma} and \emph{GraphPad} are each extracted by the
NER model under more than 200 string variations.  \emph{BLAST}, an
algorithm for comparing primary biological sequence information, has
more than 500 variations extracted by the NER model (see
Figures~\ref{fig:limma} and~\ref{fig:BLAST} with additional examples
in \ref{sec:string_variability}). Through disambiguation, our goal is
to group different string variations of the same software entity
together. For instance, all the string variations for \emph{limma}
should be grouped under the \emph{limma} software entity, all the
string variations for \emph{MATLAB} should be grouped under the
\emph{MATLAB} software entity, and so on.

\begin{figure}
  \hrulefill\\
    R microarray package limma \\
    R package “limma” \\
    R package named limma \\
    R package limma: Linear Models for Microarray Data \\
    R package limma – \\
    R package limma package \\
    R package limma (Linear Models \\
    R package limma \\
    R package like limma \\
    R package ‘limma \\
    R package for limma \\
    R package, limma \\
    R packages limma \\
    R/Biconductor package limma \\
    R/ Bioconductor package limma \\
    R-package “limma” \\
    R-package “limma \\
    R-package ‘limma’ \\
    limma package removeBatchEffect \\
    limma package for bioconductor \\
    limma package for R \\
    limma package R \\
    limma package Bioconductor \\
    limma package (Linear Models for Microarrays) \\
    limma package (Linear Models for Microarray) \\
    Bioconductor R Package limma \\
    Bioconductor R limma package \\
    Bioconductor R package limma \\
    Linear Models for Microarray Data (limma) R \\
    Linear Models for Microarray Analysis (limma) R \\
    LIMMA package (limma) \\
    GNU R limma \\
    R “limma \\
    R ‘limma’ package \\
    R software package “limma” \\
    R packages “limma \\
    R “limma” \\
    R package (limma \\
    R package 'limma' \\
    R package "limma" \\
    R Bioconductor package ‘limma \\
    R Bioconductor package limma \\
    R Bioconductor limma package \\
    R Bioconductor limma \\
    Package limma \\
    R Bioconductor package “limma” \\
    Model for Microarray data (limma) R \par
    \hrulefill\par
  \caption[Examples of string variability for the software
  `limma']{Examples of string variability for the software `limma'
    (\url{https://www.bioconductor.org/packages/limma}), as extracted
    by the NER model from the PMC-OA corpus}
  \label{fig:limma}
\end{figure}

\begin{figure}
  \hrulefill\\
    BLAST engine  \\
    BLAST output  \\
    BLAST package  \\
    BLAST)  \\
    BLAST Whole Genome  \\
    BLAST) search  \\
    BLASTING  \\
    BLASTs  \\
    BLaST  \\
    Blast  \\
    Blast -m  \\
    Blast Tool  \\
    Blast search  \\
    BLAST+  \\
    BLAST Website  \\
    BLAST Tools  \\
    BLAST Tool  \\
    (BLAST  \\
    (BLAST)  \\
    BLAST  \\
    BLAST (BLAST  \\
    BLAST (National  \\
    BLAST +  \\
    BLAST + tools  \\
    BLAST -P  \\
    BLAST Genome search  \\
    BLAST SEARCH  \\
    BLAST Search  \\
    BLAST Search tool  \\
    BLAST Searching  \\
    BLAST Sequence Similarity Search  \\
    BLAST TEXT  \\
    Blast searches  \\
    Blast tools  \\
    BLAST Searches  \\
    Blast+ package  \\
    blast sequence similarity search  \\
    blast similarity search  \par
    \hrulefill\par
  \caption[Examples of string variability for the software
  `BLAST']{Examples of string variability for the software `BLAST'
    (\url{https://scicrunch.org/browse/resources/SCR_008419}), as
    extracted by the NER model from the PMC-OA corpus}
  \label{fig:BLAST}
\end{figure}

Starting with a set of papers (in our case the PMC-OA \textsl{comm} corpus), we run
the NER algorithm and obtain the set $S$ of all software strings
extracted from the corpus. We build a similarity matrix $M$ containing
similarity scores between pairs of strings in $S$. Then we break $M$ down into distinct connected components, and run DBSCAN \cite{ester1996density}
on each. From each connected component, we obtain clusters of strings corresponding to well-defined
software entities.  The final list of clusters is the union of all the clusters from each of the connected components in $M$. To obtain $M$, we use a combination of
Keywords-based synonym generation, SciCrunch synonyms retrieval and
Jaro-Winkler string similarity scores. We go over our methodology in
more detail below.

\subsubsection{Keywords-based Synonym Generation}
\label{sec:keywords-synonym-generation}

We query PyPi~\cite{PyPi}, CRAN~\cite{CRAN} and
Bioconductor~\cite{Gentleman2004} indices, which contain lists of
software packages available on each of these platforms. For each
software entry in each index, we look for entries in the list of all
plain-text software mentions that contain the match and keywords
relevant to that index. For example, once we find \emph{limma} as an
entry in the Bioconductor package, we look for other plain-text
software mentions in our original list that contain the word
\emph{limma} and keywords relevant to the Bioconductor index, such
as \emph{limma R package, R package limma}, etc. We consider pairs
retrieved this way (e.g. \emph{limma} and \emph{limma R package}
and \emph{limma} and \emph{R package limma}) to be high-confidence
synonyms.

\begin{table}
  \centering
  \caption{Keywords used for synonym generation}
  \label{tab:disambiguation_keywords}
  \begin{tabularx}{\columnwidth}{lX}
    \toprule
    Index & Keywords Used\\
    \midrule
    PyPI & ['python, 'Python', 'API'] \\
    CRAN & ['R', 'r', 'package', 'Package', 'R-package', 'R-Package', 'r-package']\\
    Bioconductor & ['R', 'r', 'package', 'Package', 'R-package', 'R-Package', 'r-package', 'bioconductor', 'Bioconductor']\\
    \bottomrule
  \end{tabularx}
\end{table}

\subsubsection{SciCrunch Synonyms Retrieval}
\label{sec:scicrunch-synonyms-retrieval}

We query the SciCrunch API~\cite{SciCrunchAPI} for each of the mentions in
our corpus. When queried for a particular entry, the SciCrunch API has
the option of returning different variations under which that
software is known. This includes mappings between acronyms and
full names (eg \emph{SPSS} and \emph{Statistical Package for the
  Social Sciences}, but also mappings between different naming
variations, (e.\,g.\ \emph{BLASTN}, \emph{NCBI BLASTN},
\emph{Nucleotide BLAST}, \emph{Standard Nucleotide BLAST}). This
allows us to connect different naming variations of the same software
together at scale. 

\begin{table}
  \centering
  \caption{Examples of software synonym pairs retrieved through the
    SciCrunch API} 
  \label{tab:scicrunch_synoynms_examples}
  \begin{tabularx}{\columnwidth}{lXS[table-format=1]}
    \toprule
    Software Mention & Synonym \\
    \midrule
    SPSS & Statistical Package for the Social Sciences \\
    SPSS & IBM SPSS \\
    SPSS & IBM SPSS Statistics: International Business Machines SPSS
           Statistics \\
    BLASTN & Standard Nucleotide BLAST \\
    BLASTN & Nucleotide BLAST \\
    BLASTN & NCBI BLASTN \\
    BLASTN & BLASTn \\
    \bottomrule
  \end{tabularx}
\end{table}

\subsubsection{String Similarity Algorithms}
\label{sec:string-similarity-algos}

We use the Jaro-Winkler string similarity algorithm~\cite{Winkler1991}
to generate similarity scores between all pairs of software mentions
from the list of all plain-text software mentions extracted. Examples
are shown in Table~\ref{tab:keywords_synonym_generation}. We use these
scores in the similarity matrix for pairs of software mentions that we
didn't retrieve through keywords-based synonyms generation or through
the SciCrunch API.

\begin{table}
  \centering
  \caption[Examples of software synonym pairs retrieved through
  keywords synonym generation]{Examples of software synonym pairs
    retrieved through Keywords synonym generation. The synonyms are
    strings extracted by the NER algorithm}
  \label{tab:keywords_synonym_generation}
  \begin{tabular}{ll}
    \toprule Software Mention & Synonym \\
    \midrule
    scikit-learn & scikit-learn python package \\
    scikit-learn & scikit-learn python library \\
    scikit-learn & scikit-learn python\\
    scikit-learn & scikit-learn library for Python \\
    scikit-learn & scikit-learn Python package2223 \\
    scikit-learn & scikit-learn Python package for  \\
    scikit-learn & scikit-learn Python package \\
    \bottomrule
  \end{tabular}
\end{table}

\subsubsection{Clustering}
 
After we generate all the synonym pairs, we build our similarity
matrix $M$. We consider synonyms retrieved through keywords-based
synonym generation process to be high-confidence synonyms to the
original entries. Hence, we add them to our similarity matrix with the
confidence of $0.99$. We are really confident in the synonym pairs
retrieved from SciCrunch, so we give them the confidence of $1$. For
all other pairs of strings, we use the Jaro-Winkler string
similarity~\cite{Winkler1991} values in our matrix. We only use pairs
of synonyms with a confidence $\geq 0.97$. We make this choice so that
the largest connected component can fit into working memory and also
to increase the confidence in the pairs of synonyms we consider. A
drawback of this is that we will lose a number of pairs of
synonyms. Future work could include finding ways to overcome this 
limitation. Our similarity matrix is described by:
\begin{equation}
  \label{eq:Mij}
    M_{ij} = \begin{cases}
        1, & \quad (i, j) \in S_{2}\\
        0.99, & \quad (i, j) \in S_{1}-S_2\\
        m_{ij}, &\quad \text{otherwise, and }
        m_{ij} \geq 0.97 
        \end{cases},  
\end{equation}
where $S_1$ is the set of pairs based on keywords-based generation,
$S_2$ is the set of pairs obtained from SciCrunch, and $m_{ij}$ is
Jaro-Winkler distance. We perform additional post-processing of M in order to increase data quality for clustering.  Steps we include are: assigning confidences of 1 for pairs of synonyms that are equal when stripped of digits, punctuation or copyright characters, or that have more than one word token and are equal, lowercase-insensitive. We also remove pairs that include broad terms that tend to have a lot of synonyms, such as 'R package', 'r package', or 'interface'. Full details can be found in our code. 

Once we have built our similarity matrix $M$ and post-processed it, we extract the connected
components $C_1$, $C_2$, \dots, $C_n$ of $M$ and then run the DBSCAN \cite{ester1996density}
algorithm on each similarity matrix $M^{(n)}$ corresponding to the
component number $n$. Computing the connected components helps us to
reduce the size of the matrix we have to run the clustering algorithm
on. For each $M^{(n)}$ we compute a distance matrix $D^{(n)}$ such that
\begin{equation}
  \label{eq:Dij}
  D_{ij}^{(n)} = 1 - M_{ij}^{(n)}.
\end{equation}

Using DBSCAN we obtain a set of clusters corresponding to each
connected component in our graph. For each cluster, we select the
software mention with the highest frequency in our corpus as
the cluster name.  Empirically, we noticed that this string variation is the most
likely to be the real name of the software entity. This also
corresponds to the intuitive idea of the ``true software mention''.

Our algorithm is presented as Algorithm~\ref{alg:clustering}.  

\begin{algorithm}
  \caption{Clustering and naming}
  \label{alg:clustering}
  \textbf{Input} $\text{{distance matrices $D_{ij}^{(n)}$}}$ \\
  \textbf{Output} $\text{all\_clusters} \leftarrow \{\}$ \\
  \begin{algorithmic}
    \FORALL {connected components $C_n$ of $M$ with distance matrices
      $D_{ij}^{(n)}$} 
    \STATE $\text{Cluster}_1^{(n)}$,
    $\text{Cluster}_2^{(n)}$, \ldots $\gets\DBSCAN(D^{(n)})$
    \FORALL {clusters $\text{Cluster}_i^{(n)}$ from component $C_n$}
       \STATE Select vertex
       $s_i^{(n)}\in\text{Vertices}(\text{Cluster}_i^{(n)})$ with the
       highest frequency on PMC-OA \textsl{comm}
       \STATE $\text{Name}(\text{Cluster}_i^{(n)})\gets s_i^{(n)}$
       \STATE $all\_clusters \leftarrow all\_clusters \cup \text{Cluster}_i^{(n)}$
    \ENDFOR
    \ENDFOR
  \end{algorithmic}
\end{algorithm}

We use the textdistance python package~\cite{textdistance} to compute
the Jaro-Winkler similarity scores. We use a CSR sparse matrix format
for the similarity matrix and use the connected\_components module
from scipy.sparse.csgraph ~\cite{SciPy} to obtain the connected
components. We use the DBSCAN implementation from
sklearn.cluster~\cite{scikit-learn}.

\subsection{Software Mentions Linking}
\label{sec:linking}

Once we cluster different software string variations together,
we also want to link them to the corresponding URLs. For each software
mention in our corpus, we do an exact match search in the PyPI~\cite{PyPi}, CRAN~\cite{CRAN}
and Bioconductor~\cite{Gentleman2004} indices, as well as GitHub~\cite{GitHub} and SciCrunch~\cite{SciCrunch} APIs . For PyPI, CRAN, Bioconductor and SciCrunch, we also query
individual URL pages for matches we find. Combined, we obtain links,
as well as additional metadata, for a number of software
mentions. Because the format and type of metadata available varies
across repositories, we normalize the formats to a common schema
between repositories. The final metadata fields we retrieve across
repositories and normalizing to contain:
\path{source}, \path{package_url}, \path{description, homepage_url},
\path{other_urls}, \path{license}, \path{github_repo},
\path{github_repo_license}, \path{exact_match}, \path{RRID},
\path{reference}. Note that not all software
mentions will have all of the metadata fields present, either because
there was no corresponding field in the database for that field, or
the entry was empty in the database. We make the schema normalization
mappings between the initial fields present in a database and the
final metadata fields available in Table~\ref{tab:normalized_metadata}
and~\ref{sec:normalization}.

We map each cluster to the link to which the cluster name is pointing
to. For instance, we map all the entries in the \emph{scikit-learn}
cluster to the \emph{scikit-learn} link. For situations where the
cluster name does not have an associated link, or a software mention
is not mapped to a cluster, we look for a an exact match link for that
particular software mention itself.

\begin{table}
  \centering
  \caption[Normalized metadata]{List of normalized metadata, together
    with definitions, for fields obtained by linking through the PyPI,
    CRAN, Bioconductor indices and SciCrunch and GitHub APIs.}
  \label{tab:normalized_metadata}
  \begin{tabularx}{\columnwidth}{lX}
    \toprule
    Field & Definition \\
    \midrule
    database & database the software is being mapped to \\
    package\_url & URL of the software in the database \\
    description & description of the software in the database \\
    github\_repository & GitHub repository for the software \\
    homepage\_url & homepage for the software (retrieved from the database)\\
    other\_URLS & list of other URLs retrieved for the software from the database\\
    reference & journal article linked to the software, identified by DOI, PMID or RRID \\
    scicrunch\_synonyms & synonyms for software retrieved from SciCrunch\\
    resource\_type & resource type according to SciCrunch\\
    RRID & RRID for the software retrieved from SciCrunch\\
    \bottomrule
  \end{tabularx}
\end{table}

\subsection{Dataset Curation}
\subsubsection{Motivation}

The NER model has an F1 score of 0.922, with a precision of 0.9063 and
a recall value of 0.9385. These metrics are calculated on the SoftCite
dataset~\cite{Du2021}, which has been used for training the NER
model. Since the PMC-OA corpus distribution is different from the one
of the SoftCite Dataset, we engaged our in-house team of bio-curators
to first evaluate a sample of the top 1000 plain text software
mentions (by means of frequency) extracted from the PMC-OA  \textsl{comm} subset.
Based on this evaluation, we concluded that the Precision@1000 of
the NER model ``in the wild'', on the PMC-OA corpus is 79.5\%. After
this assessment, and in order to eliminate as much noise from the
dataset as possible, we engaged our bio-curator team to curate another
set of top \num{9000} mentions (by frequency), thereby generating a
dataset of \num{10000} curated mentions in total. Each of the
\num{1000} mentions was annotated by a single curator and checked by a
second curator; the \num{9000} mentions were curated by one curator
each. After excluding terms that our biocurators mark as
\emph{non-software}, we are left with \num{6966}~mentions
that meet our definition of software (for definitions
see~\ref{sec:curation}), as shown in
Tables~\ref{tab:curation_coverage}, \ref{tab:evaluation_10k},
and~\ref{tab:evaluation}. We curate all three corpora: \textsl{comm},
\emph{non\_comm} and the \emph{Publishers' collection} datasets to
exclude mentions marked by curators as \emph{non-software}. We also
make the raw datasets available, together with the curators' labeling,
and comments.

\begin{table}
  \centering
  \caption[Metrics of curated datasets]
  {Metrics of curated datasets. The table shows the coverage of the
    \num{10000} curated mentions in the PMC-OA and CZ Publishers'
    Collection datasets. The curated mentions are the top 10k mentions
    in terms of frequency on the PMC-OA comm dataset}
  \label{tab:curation_coverage}
  \begin{tabular}{lS[table-format=2.2e1]S[table-format=2.3]}
    \toprule Corpus & \multicolumn{2}{p{11em}}{\centering Mentions covered by the 10k
                      set}\\
    \cmidrule{2-3}
    & {Number} & {\% of the set}\\
    \midrule
    PMC-OA comm & 9.03e6 & 61.158 \\
    PMC-OA non\_comm & 2.67e6 & 58.923 \\
    CZ publishers\_collection & 20.11e6 & 41.766\\
    \bottomrule
  \end{tabular}
\end{table}

\subsubsection{Curation Guidelines \& Evaluation}

Our software definitions, are broadly based on the definitions used
for the SoftCite dataset~\cite{Du2021}, on which the NER model has
been trained. We start with five different categories:
\textit{software, algorithm, database, web platform, hardware} and
\textit{other} and ask our curators to annotate the top
\num{1000}~most frequent mentions from our corpus with these
categories. To facilitate the curation process, we provide for each
software mention five different sentences extracted from the articles;
where the sentences do not provide sufficient information about the
software, the curators are asked to find background information
through Google searches. The curation of the top \num{1000} mentions
allowed us to (a)~understand the variety of mentions that are present
in the dataset (see Table~\ref{tab:evaluation}) and (b)~identify
additional examples for each category (in particular, borderline cases) and refine the curation
guidelines further.

We consider mentions in the \textit{software} and \textit{algorithm}
categories to be true \emph{software}, whereas mentions in the
\textit{database, hardware, web platform} and \textit{other}
categories are considered as \emph{non-software}. Based on this human
evaluation, 79.5\% of \num{1000}~mentions are evaluated as
\emph{software}, and 16.5\% as \emph{non-software}. For the remaining
4\% of mentions, a distinction between software and non-software
could not be made; this is either because a software name or acronym
could be pointing to two different entities (labeled as
\textit{unclear}), or because the mentions extracted by the NER model
were random symbols. For the evaluation of the \num{9000} mentions, we
only ask our curators to distinguish between the main
\emph{software/non-software} categories. We provide more detailed
curation guidelines in~\ref{sec:curation}. In the final curated dataset of 10000 
software mentions, 69.66\% of mentions are evaluated as
\emph{software}, 21.55\% as \emph{non-software}, and for 8.79\% an evaluation 
could not be made.

\begin{table}
  \centering
  \caption{Expert evaluation of most frequent mentions extracted by the NER model from the PMC-OA \textsl{comm} dataset}
  \label{tab:evaluation_10k}
  \begin{tabular}{l*{2}{S[table-format=2.2]}}
    \toprule
    Category & {Precision@1k, \%} & {Precision@10k, \%}\\
    \midrule
    software & 79.5 & 69.66\\
    not-software & 16.5 &  21.55\\
    unclear &4 & 8.79 \\
    \bottomrule
  \end{tabular}
\end{table}

\subsubsection{Inter-Annotator Agreement}

We compute the Inter-Annotator Agreement (IAA) value by looking at
curator evaluations on a random sample of 100 mentions from the larger
sample of \num{1000}~(Table~\ref{tab:evaluation}. With four curators
per mention, we obtain a Fleiss Kappa IAA value~\cite{Gwet2014} of
0.639 when only the two main categories (software/non-software) are
considered, and 0.504 when the more specific five subcategories
(software, algorithm, database, web platform, hardware, other) are
labeled. We also compute the Krippendorff Alpha IAA~\cite{Gwet2014}
values, which are similar: 0.686 for the two main categories and 0.523
for five subcategories (Table~\ref{tab:IAA}). Since we are interested
in the binary (software/non-software) differentiation, our IAA values
of concern are around 0.639/0.686. We want to note that the relatively
low IAA value demonstrates that this is a challenging task even for
our expert curator team. The fact that the IAA value is relatively low
should be interpreted as a signal for how hard it is to distinguish in
some cases whether a mention is a true software or not. Specifically,
it is difficult to differentiate between `software' and `algorithm' as
these terms are often used interchangeable by authors describing their
tools: the tool may be called ``software'' on the tool’s website but
described as an ``algorithm'' in an article. Similarly, some online
tools consist of a database and software of the same name, so there
may be disagreement how curators annotate the software mention. It is
also important to note that the curation relied on both the context
given by 5 example sentences and background searches. The use of different
sources of information can result in different annotations; this is particularly true for software
mentions that consist of acronyms, have ambiguous names or share part of their name with
other software and/or non-software resources.

\begin{table}
  \centering
  \caption{Expert evaluation of \num{1000} most frequent mentions extracted by the NER model}
  \label{tab:evaluation}
  \begin{tabular}{llS[table-format=3]S[table-format=2.1]}
    \toprule
    Main Category & Subcategory & {Counts} & {\%}\\
    \midrule
    Software & Software & 699 & 69.9 \\
    Software & Algorithm & 95 & 9.5 \\
    Non-Software & Database & 40 & 4.0 \\
    Non-Software & Hardware & 9 & 9 \\
    Non-Software & Web platform & 13 & 1.3 \\
    Non-Software & Other & 105 & 10.5 \\
    Errors & Unclear & 39 & 3.9 \\
    \bottomrule
  \end{tabular}
\end{table}

\begin{table}
  \centering
  \caption{IAA curation values for the expert evaluation of the most frequent mentions extraced by the NER model from the PMC-OA \textsl{comm} dataset}
  \label{tab:IAA}
  \begin{tabular}{l*{2}{S[table-format=1.3]}}
    \toprule
    Parameter & {Five categories} & {Two categories}\\
    \midrule
    Fleiss $\kappa$ & 0.504 & 0.639 \\
    Krippendorff $\alpha$ & 0.523 & 0.686 \\
    \bottomrule
  \end{tabular}
\end{table}

\section{Evaluation \& Results}

\subsection{Final Dataset}

The statistics of software mentions extracted from the OA-PMC \textsl{comm} is
shown in Table~\ref{tab:oapmcstats}.  The results of disambiguation
and linking are shown in Table~\ref{tab:oapmc_disambiguation}.  As
seen from this table, for about \num{390000}~mentions we could not
generate synonyms or we discarded them during post-processing.  By analyzing some of these mentions, we
hypothesize that they are likely to be noise or false positives.   For
about \num{720000}~mentions we were able to generate synonyms. Again,
about \num{400000}~of the latter were too ambiguous to
disambiguate. This is likely because the software mentions did not
have \emph{significant} synonyms (with a confidence of $\geq$ 0.97
during the disambiguation phase).  The remaining \num{320000}~mentions
were mapped into \num{97600}~unique software entities through the
disambiguation algorithm.  This covers 78\% of all links in the
dataset.  Lastly, for about \num{185000}~mentions we were able to
obtain links to GitHub, PyPI, CRAN, Bioconductor, or SciCrunch. This
covers about 55\% of all software-paper links.

\begin{table*}
  \centering
  \caption{OA PMC \textsl{comm} disambiguation and linking}
  \label{tab:oapmc_disambiguation}
  \centering
  \begin{tabular}{lS[table-format=7]S[table-format=3.2]S[table-format=7]S[table-format=3.2]l}
    \toprule
    Category
    & \multicolumn{2}{c}{Mentions} &  \multicolumn{2}{c}{Paper-software
                                   links} &
                                   Notes\\
    \cmidrule{2-5}
    & {Approx. \#} & {\%} & {Approx. \#} & {\%}\\
    \midrule
     undisambiguated & 393057 & 35 & 731529 & 8.95 &  no significant synonyms \\
     undisambiguated & 404493 & 36.11 & 1050017 & 12.85 &  no output from clustering \\
     disambiguated & 323561 & 28.88 & 6384430 & 78.18 &  \num{97600} unique software entities \\\addlinespace
     disambiguated \emph{and} linked & 185427 & 16.55 & 4555476 & 55.78&
    \\\addlinespace 
    total unique mentions & 1120111 & 100 & 8165976 & 100 &    \\
    \bottomrule
  \end{tabular}
\end{table*}

\subsection{Disambiguation}

Perhaps the biggest challenge in disambiguation is that we don't have curated labels to learn from.  Our methodology, based on DBSCAN \cite{ester1996density}, is
unsupervised. We engaged our biocurators for evaluation of the
results. 

 We created a set of \num{5884}~pairs of generated synonyms coming from \num{104}~unique
software mentions for manual curation. This dataset was generated from an initial iteration of the disambiguation algorithm. We asked our curation team to label each synonym pair with one of the following
categories:
\begin{description}
    \item[Exact:]  synonym is an exact match of the assigned Software mention 
    \item[Narrow:]  synonym is a child term
    \item[Not Synonym:]  synonym is an unrelated software or other term
\end{description}

The \textit{Exact} category includes all mentions that can
unambiguously be assigned to the software mention, including partial
terms, typos, variations on spelling (upper/lower case, space, hyphen
etc), and acronyms. Reference numbers in the article are often
erroneously added to the software synonym; these are also labeled as
\textit{Exact}.  Versions of software are annotated as
\textit{Narrow}. Partial versions are also included here, for example
``Autodock Vina1'' for ``Autodock Vina1.1.2''. It is worth noting that
synonyms representing software versions can be challenging to
distinguish from synonyms which include the reference number
(e.g. ``Autodock Vina19'' where 19 is the reference number in the
article mentioning the software).  Synonyms of software tools that
have common names pose another challenge for this task: for example,
it is time-consuming to work out whether ``ClusterM'' is a true
synonym for a version of the software tool ``Cluster'' or represents a
completely distinct software called ClusterM.  We considered pairs that
fall into either of \textit{Exact} or \textit{Narrow} categories to be
true synonyms. Composition of this curated labels is under (Table~\ref{tab:disambiguation_evaluation}). 

We used this dataset and the labels assigned by our curators to validate our clustering technique, by computing the Precision, Recall and the F1 score for generated pairs of synonyms.  We chose the hyperparamaters that gave the highest metrics on this curated dataset. Our best model had an F1 score of 0.704, with a  Precision of 0.954 and Recall of 0.558.

\begin{table}
  \centering
  \caption[Disambiguation Evaluation]{Disambiguation
    Evaluation. Description of the dataset used for evaluating disambiguation. 
    Labels were assigned by our curation team on 5886 pairs of generated synonyms coming
    from \num{103}~unique software mentions}
  \label{tab:disambiguation_evaluation}
  \begin{tabular}{lS[table-format=4]S[table-format=[2.3]}
    \toprule
    Synonym Pair Label & {Count} & {\%}\\
    \midrule
    Correct - Exact & 3147 & 53.465 \\
    Correct - Narrow & 1094 & 18.586 \\
    Incorrect & 668 & 13.484 \\
    Unclear & 45 & 0.908 \\
    Not software & 930 & 15.805 \\
    \bottomrule
  \end{tabular}
\end{table}

We offer some examples of disambiguated terms obtained through DBSCAN clustering for the \textbf{scikit-learn}, \textbf{ImageJ} and \textbf{SPSS} software entities in Figure~\ref{fig:results-scikit-learn}, Figure~\ref{fig:results-ImageJ} and Figure~\ref{fig:results-SPSS}.

\begin{figure}
   \textbf{Cluster: `scikit-learn`} \\
   \hrulefill \\
   Scikit-Learn \\
   scikit-learn Python \\
   scikit-learn \\
   Scikit learn \\
   sklearn \\
   scikit-learn Python library \\
   Scikit-learn \\
   Sklearn \\
   Python scikit-learn \\
   scikit learn \\
   scikit‐learn \\
   Scikit‐learn \\
   Scikit-Learn Python \\
   scikit-learn81 \\
   SciKit-Learn \\
   sklearn Python \\
   SkLearn \\
   scikit -learn \\
   Python Scikit-Learn \\
   Scikit-Learn® \\
   Python sklearn library \\
   scikit.learn\\
   Scikits-learn \\
   Sci-kit-learn \\
   scikit-learn Python3 package\\
   Sci-Kit learn\\
   sklearn0 \\
   Python Scikit Learn\\
   Scikit-Learn Python Library \\
   Python scikit\_learn\\
   2scikit-learn \\
   SCikit-learn\\
   SCIKIT-learn\\
   Scikit–Learn\\
   Python Sci-kit learn \\
   Python package Scikit-learn\\
   scikit-Learn\\
   Scikit-Learn Python package\\
   scikit-learn Python Library\\
   sklearn python package\\
   sci-kit learn Python package\\
   Python scikit learn package\\
   skLearn \\
   Scikit-learn python\\
   scikitslearn\\
   scikit-learn Python package for \\
   Python sci-kit learn\\
   Scikit–Learn Python \\
   scikit - learn Python library\\
   ....\par
    \hrulefill\par
  \caption[Sample of Disambiguation results for the software entity  \textbf{scikit-learn}]{Sample of Disambiguation results for the software entity \textbf{scikit-learn}. There are a total of 124 unique string variations extracted by the NER model that are mapped to the \textbf{scikit-learn} software entity through disambiguation.}
  \label{fig:results-scikit-learn}
\end{figure}

\begin{figure}
   \textbf{Cluster: `ImageJ`} \\
   \hrulefill \\
   ImageJ \\
   Image J \\
   image J \\
   Image-J \\
   Image J1 \\ 
   Image J2 \\
   Image J© \\
   Image/J® \\
   image/J® \\
   image-j\\
   IMAGE-J\\
   ImageJ (Image Processing and Analysis in Java \\ 
   Image-j\\
   Image J) \\
   IMAGE j/ \\
   IMageJ \\
   imageJ1\\
   Image -J\\
   ImageJ®1 \\
   ImageJ, Image Processing and Analysis in Java\\
   ImageJ (Image Processing and Analysis in Java)\\
   ImageJ Image Processing and Analysis in Java \\
   ImageJ-145 \\
   image J1 \\
   IMageJ1 \\
   imageJ64\\
   ImageJ -\\
   ImageJ)\\
   IMAGEJ® \\
   Image/J\\
   Image J-Image Processing and Analysis in Java\\
   Image J- Image Processing and Analysis in Java'\\
   ....\par
    \hrulefill\par
  \caption[Sample of Disambiguation results for the software entity \textbf{ImageJ}]{Sample of Disambiguation results for the software entity \textbf{ImageJ}. There are a total of 178 unique string variations extracted by the NER model that are mapped to the \textbf{ImageJ} software entity through disambiguation.}
  \label{fig:results-ImageJ}
\end{figure}

\begin{figure}
   \textbf{Cluster: `SPSS`} \\
   \hrulefill \\
   Statistical Package for Social Sciences (IBM – SPSS)\\ 
   Statistical Package for the Social Sciences (SPSS) Package\\
   (Statistical Package for Social Science)\\
   (Statistical Package for the Social Sciences\\
   SPSS®® Statistics\\
   Statistical Package for the Social Sciences [(SPSS\\ SPSS\\
   Statistical Package for the Social Sciences (SPSS -19\\
   Statistical Package for Social science for Windows (SPSS)\\
   SPSS -24\\
   SPSS27\\
   Statistical Package for Social Sciences (SPSSR-25\\
   Statistical Package for Social Sciences (IM SPSS\\
   SPSS® Statistics®\\
   Statistical Package for Social Sciences-24\\
   Statistics Package for Social Sciences\\
   Statistical Package for the Social Science\\
   SPSS* Statistics\\
   Statistical Package for Social Sciences [SPSSTM]\\
   SPSS]\\
   Statistiscal Package for the Social Sciences \\
   PSS Statistics\\
   Statistical Package for the Social Sciences for Mac\\
   Statistical Package for Social Sciences (SPSS-2\\
   Statistical Package for the Social Sciences (SPSS-23\\
   Statistical Package for the Social Sciences Software  (SPSS\\
   SPSS15 \\
   Statistical Package for Social Sciences (SPSS-21\\
   spss20\\
   Statistical Package for the Social Science program (SPSS\\
   statistical package for the social sciences (SSPS)\\
   Statistical Package for the Social Sciences Software\\
   Statistical Pack for the Social Science (SPSS)\\
   Statistical Package for Social Science (SPSS) (SPSS\\
   Statistical Package for the Social Sciences for Mac (SPSS‐22\\
   statistical package for Social Science (SPSS PC\\
   Statistical Package for the Social Sciences (SPSS) (PASW\\
   SPSS)\\
   SSPS Statistics\\
   IBM statistical package for the social sciences (SPSS)\\
   Statistical Package for the Social Sciences, SPSS)\\
   SPSS tatistics\\
   ....\par
    \hrulefill\par
  \caption[Sample of Disambiguation results for the software entity \textbf{SPSS}]{Sample of Disambiguation results for the software entity \textbf{SPSS}. There are a total of 1361 unique string variations mapped to the \textbf{SPSS} cluster.}
  \label{fig:results-SPSS}
\end{figure}

\subsection{Linking}

As with disambiguation, we don't have any labeled data to evaluate our linking, so we engage our team of biocurators for feedback. We provide our biocurators with 50~software mentions, together with the
generated links. Based on biocurator feedback, 54\% the generated
links are correct, 6\% are incorrect, and for 40\% it is unclear
whether the link is correct. Most notably, 39/40 mentions for which
the link is unclear are retrieved from GitHub and only one is linked
through PyPI, Bioconductor, CRAN and SciCrunch. This happens because
GitHub is a resource that is not curated, so having an exact match on
a software name in GitHub does not guarantee that it will be linking
to the actual software.  Anyone can upload software in GitHub and name it as they wish.  A number of repositories we evaluated were empty, or contained too little information to be able to decide for sure if the linking was correct. When we consider the evaluation only of links
retrieved through PyPI, Bioconductor, CRAN and SciCrunch, the accuracy
improves considerably: 93.33\% of links are correct and 0.6\% are
incorrect, 0\% unclear.  This suggests that linking software through
these four repositories through an exact match is likely to give
correct links. We note, however, that the sample size for this
evaluation is quite small, and it is possible that results might
change with a larger sample.

\begin{table}
  \centering
  \caption[Linking Coverage]{Linking Coverage}
  \label{tab:linking_coverage}
  \begin{tabular}{lS[table-format=3]S[table-format=3.3]}
    \toprule
    Repository & {Number of linked mentions} & {\%}\\
    \midrule
    GitHub API & 155506 & 64.39 \\
    SciCrunch API & 43817 & 18.14 \\
    CRAN & 20202 & 8.36 \\
    PyPI & 14154 & 5.86 \\
    Bioconductor & 7801 & 3.23 \\
    \bottomrule
  \end{tabular}
\end{table}

\begin{table}
  \centering
  \caption[Linking Evaluation]{Linking Evaluation on a sample dataset of
    50 generated links to PyPI, CRAN, Bioconductor, GitHub and
    SciCrunch}
  \label{tab:linking_metrics}
  \begin{tabular}{l*{3}{S[table-format=2]}S[table-format=2.2]}
    \toprule
    Link label & \multicolumn{2}{c}{All links} &
                                                 \multicolumn{2}{c}{Excluding
                                                 GitHub}\\
    \cmidrule{2-5}
               & {Number} & {\%} & {Number} & {\%} \\
    \midrule
    correct & 27 & 54 & 14 & 93.33\\
    unclear & 20 & 40 & 1 & 6.66\\
    incorrect & 3 & 6 & 0 & 0\\
    \bottomrule
  \end{tabular}
\end{table}

\begin{table}
  \centering
  \caption[Metadata Fields Statistics for linked mentions]{Metadata
    Fields Statistics for the mentions linked to a repository.
    Synonyms retrieved through diasmbiguation that might link to the
    same entry are not counted}
  \label{tab:metadata_stats}
  \begin{tabular}{lS[table-format=6]S[table-format=3.3]}
    \toprule
    Normalized metadata field & {Count} & {\%}\\
    \midrule
    package\_url & 149015 & 100\\
    github\_repo & 143834 & 96.52\\
    description & 116071 & 77.89\\
    homepage\_url & 36306 & 24.36\\
    github\_repo\_license & 39464 & 26.48\\
    reference & 22134 & 14.85\\
    RRID & 18766 & 12.59\\
    other\_urls & 18766	 & 12.59\\
    license & 13485 & 9.04\\
    \bottomrule
  \end{tabular}
\end{table}

\section{Code \& Data Availability}

We make the dataset, as well as all intermediate files available at
\url{https://doi.org/10.5061/dryad.6wwpzgn2c} under a CC0
license~\cite{Dataset}. All the code used for extraction,
disambiguation and linking, as well as instructions on how to
reproduce the results and some starter code is available at a GitHub
repository \url{https://github.com/chanzuckerberg/software-mentions}
under the MIT license with the permanent snapshot at~\cite{Code}.

\section{Discussion}
\label{sec:discussion}

The dataset we make available is, to our knowledge, the largest
dataset of software mentions in the scientific literature currently
available.  In this section, we go over some of the lessons learned in
building a dataset this large, as well as limitations of current
work. We offer a view on opportunities for further work in the
\textit{Next Steps} section.

\subsection{Extraction}
\label{sec:discussion_extraction}

An important feature of our extraction is the maximal preservation of
the context of the software mention.  We keep the sentence from which
the mention was extracted as well as the reference to the unit
(section, caption, abstract) from which it was extracted.  This helped
in the curation.  This may facilitate more deep reprocessing of the
dataset the future, including better disambiguation and linking.  This
feature became available because we had the access to the structure of
the papers, using their XML representation.  Many other studies, for
example, \cite{Lopez2021dataset}, process the text extracted from PDF,
which does not show the structure.  The current tendency of the
publishers to keep the archival copies of the papers in the XML format
is probably one of the best things that happened to the field of data mining from scholarly documents.

\subsection{Software Name Variability in the Literature} 

The goal of building a comprehensive dataset of software mentions in
the literature is to be able to assess software impact through
informal citations. One of the main challenges is the variability
under which a software can be mentioned, or 'informally cited' in a
paper. Some of this variability is to be expected, such as differences
between using the fullname or the acronym of a software (eg \textit{SPSS} and \textit{Statistical Package for Social Sciences}), authors using
various software name variations (eg \textit{SPSS Statistics}, \textit{IBM SPSS}) or typos.

But probably the biggest string variability comes from how the NER model extracts software from text. For instance, the NER model might extract the
following mentions: \textit{R package limma}, \textit{limma R package}, \textit{Limma}, all pointing out to the same software. 

Lastly, there is variability coming from typos that are occuring
either due to the authors or because of XML parsing. If we want to
truly get the impact of a particular piece of software, we need to be
able to map all of these software naming variations to the same
software entity.

\subsection{Disambiguation} 

In order to map software variations to the same software entity, we
built a software disambiguation model. We describe the model, based
largely on string similarity algorithms and clustering techniques,
under Section~\ref{sec:disamb}. Note that through this model, we are
only able to disambiguate 28.88\% mentions that constitute 78.18\% of
paper-software links. The other mentions either have no significant
synonyms or are too ambiguous to disambiguate
(Table~\ref{tab:oapmc_disambiguation}). We propose this as a baseline
model. Because the raw data is available, more sophisticated
disambiguation algorithms can be employed. We want to note that the
main challenge in disambiguating mentions in a dataset this large is
the lack of labels, which means that the disambiguation tasks would
have to be unsupervised. This poses challenges in terms of
evaluation. We evaluated our disambiguation method using our
biocuration team, which evaluated \num{5884}~synonym
pairs that we subsequently used to validate our model. The other challenge in using unsupervised methods, such as clustering, on a dataset this large, is the generation of distance
matrices that would fit into working memory.

We also want to note that we experimented with using transformer-based
models to generate embeddings for software mentions and compute
similarity scores between pairs of strings by using the dot product or
cosine similarity. We obtained
unsatisfactory results creating similarity scores in this way. Pairs
of strings that are semantically different than each other ended up
having similar embeddings. This makes sense, as transformer-based
models compute representational encodings for sequences based on
context. It is possible that packages that are similar in some way,
like \textit{sklearn} and \textit{scipy}, appear in similar contexts,
and hence, will have similar embeddings. We hypothesize that the level
of context around how a particular software is mentioned in a research
paper is not granular enough to allow the computation of an embedding
specific enough to differentiate that software from other similar, but
different softwares. Basically, the language (and hence context) which
authors use to mention \textit{sklearn} is likely to be very similar
to the language authors use to mention \textit{scipy}. It makes sense
that their embeddings will be close in $n$-dimensional space. 
We hypothesize that being able to use the context around a mention would, 
however, be useful for ambiguous cases when two different software can have the same name, 
or acronyms. This would be an interesting area of further work.

\subsection{Use cases}
\label{sec:use_cases}

As a philanthropic organization supporting and building technology to
serve the computational needs of biomedical scientists, we rely on
data on scientific software to inform our programmatic
activities~\cite{Sofroniew2019, CZIScience} and technology
strategy~\cite{CZITech}.  In this section we discuss a couple of ways
the dataset has already been used in our work at the Chan Zuckerberg
Initiative prior to the public release.

\subsubsection{Trends in imaging}
\label{sec:imaging}

The Chan Zuckerberg Initiative considers imaging to be one of the key
technologies in our effort to develop science and technologies to
measure human biology in action \cite{CZI_Bioimaging, CZI10Year}.  To fund this area more efficiently and support the relevant computational tools, we needed to understand the methods used in the field and the penetration of open source ones, in particular. 

Modern biomedical imaging is significantly computational in nature. Thus we were able to use our data to identify the papers using various imaging software. We used publication data to get the time trends and the change of relative market share for both open and closed source imaging software.

\subsubsection{Penetration of single-cell methods in biomedicine}
\label{sec:single-cell}

Single-cell methods are another key area of biomedical technology supported by 
CZI~\cite{CZI10Year}.  To better focus our efforts, we needed data on the penetration of single-cell methods in the clinical research literature.  The
software used for analysis of single cell data is specific, so mention
of this software in a paper can be a good marker of the usage of
single cell methods.  We were able to get a subset of papers that used
single cell methods, and to estimate the penetration of these methods
in the different biomedical subfields.

\subsubsection{Other usage}
\label{sec:other}

We would like to note that the use cases mentioned in this section
were also recognized by other people, for example, Wikidata Scholia
project \citep{NielsenF2017Scholia}.  There are many other interesting
questions, like the identifying the most influential authors, co-uses
etc.~\citep{NielsenF2017Scholia}. We hope our dataset can be used by
projects like Scholia and intend to integrate our results with
Wikidata in the future releases.

\section{Next Steps}
\label{sec:next_steps}

We believe that there are a number of exciting questions this dataset
can help answer. To start with, we can look into what the most used
software is in a particular field (e.g. \textit{neurodegeneration,
  single-cell biology, imaging}). We can go a step further and try to
understand what the software is used for. Potential ideas include
getting insights from paper or sentence topics, MESH terms,
author-provided paper keywords, or information found in the sentence
in which the software is being mentioned (note that we make this
available). We can also start looking at differences between how
software usage differs between particular fields, if any. Furthermore,
we can explore measures of impact for software, whether those are
number of informal citations (such as software mentions), or more
sophisticated models. For instance, since we are now able to connect
papers and software entities in an underlying graph, we can start
exploring with graph-based models for impact. One particularly
exciting idea would be to extend the notion of Eigenfactor, which has
been proposed measure impact for journals and authors, to software.
Other potential areas of further research include looking into
open-access policies of most used or impactful software or exploring
differences in how software usage varies over time. Last but not
least, we can look into the impact of particular pieces of software
and assess their impact.

The approach we took in this paper was to use an NER model to collect
software string variations from a corpus of papers, and then
disambiguate and link these mentions into clusters of software
entities using DBSCAN \cite{ester1996density} and a similarity matrix we built. 
Using this methodology, we were able to disambiguate about a third of the total
number of unique mentions in our dataset, which covers about 78\% of the total paper-software links. This means that we are losing information from the rest of the total number of mentions in our dataset. Future work can include improving the disambiguation algorithms to be able to cluster a larger percentage of our dataset
under particular software entities.

For linking, we used an exact match search on a software mention in
some of the most used software repositories, such as PyPI, CRAN,
Bioconductor, SciCrunch and GitHub. We want to acknowledge that there
are limitations in this approach. For instance, there can be different
software having the same acronyms, or different entities, whether
software, databases, hardware or platforms with the same
name. Moreover, in repositories that are not curated, such as GitHub,
anyone can post a piece of software under whatever name they
want. Being able to distinguish which is the true link for a piece of
context would most likely require being able to look at the context in
the text around a piece of software is being mentioned. Since we make
the sentences in which a software mention appears available, we
believe linking algorithms based on context are worth
exploring. Improving the quality of linking software mentions to
corresponding URLs in repositories or databases is an interesting area
for further research.

\section{Conclusion}
\label{sec:concl}

We created one of the largest dataset of software mentions in the literature.  For a subset of the data we used string similarity algorithms and unsupervised clustering techniques to disambiguate the software mentions into distinct software entities. We used a linking algorithm to connect the mentions to URLs in the PyPI, CRAN,
Bioconductor indices and the SciCrunch and GitHub APIs.  We make the
dataset available to the community and believe there are a number of
exciting next steps. It is our hope that this new resource helps
foster new insights about software usage and impact in the literature.

\section{Acknowledgements}
\label{sec:acknowled}

We thank James L Howison (University of Texas), Daniel Mietchen (Ronin Institute) and other reviewers for helpful comments on earlier versions of this manuscript. We also thank our team of biocurators, Alison Jee, Celina Liu, Parasvi Patel, and Ronald Wu for their hard work curating the final dataset.

\bibliography{mentions}
\bibliographystyle{unsrtnat}

\appendix

\section{Schema Normalization}
\label{sec:normalization}

In this section, we offer more details about the process of
normalizing schemas across the PyPI, CRAN and the Bioconductor
indices, as well as the SciCrunch and GitHub APIs. By querying each of
these repositories, we retrieve a number of metadata fields, which we
map to a normalized set of fields
(Table~\ref{tab:schema_normalization}).  

\begin{table*}
  \centering
  \caption[Schema Normalization]{Schema Normalization. Mappings
    between Normalized fields and corresponding fields from each resource: PyPI, CRAN, Bioconductor, SciCrunch API, GitHub API}
  \label{tab:schema_normalization}
  \begin{tabularx}{\textwidth}{*{6}{X<{\raggedright}}}
    \toprule
    Normalized Field & PyPI & CRAN & Bioconductor & SciCrunch API & GitHub API\\
    \midrule
    mapped\_\allowbreak to & PyPI package & CRAN Package & Bioconductor Package & Resource Name & best\_github\_match \\
    source & PyPI Index & CRAN Index & Bioconductor Index & SciCrunch API & GitHub API\\
    platform & PyPI & CRAN & Bioconductor & SciCrunch & GitHub\\
    package\_\allowbreak url & pypi\_\allowbreak url & CRAN Link & Bioconductor Link & Resource ID Link & github\_url\\
    description & query pypi\_\allowbreak url page & Title & Title & Description & Description\\
    homepage\_\allowbreak url & query pypi\_\allowbreak url page & query CRAN Link & query Bioconductor Link & Resource Name Link & github\_url\\
    other\_\allowbreak urls & None & None & None & Alternate URLs + Old URLs & None\\
    github\_\allowbreak repo & query pypi\_\allowbreak url page & query CRAN Link & query Bioconductor Link & SciCrunch API & github\_url\\
    github\_\allowbreak repo\_\allowbreak license & None & None & None & None & github\_repo\_license\\
    reference & None & query CRAN Link & query Bioconductor Link & Reference Link or Proper Citation & GitHub API\\
    RRID & None & None & None & Resource ID & None\\
    scicrunch\_\allowbreak synonyms & None & None & None & synonyms & None\\
    \bottomrule
  \end{tabularx}
\end{table*}

\section{Curation Guidelines}
\label{sec:curation}

\subsection*{Background}

A subset of \num{10000} software mentions was curated by our domain experts, so as
to identify the mentions that meet our definition of ``Software'' and
exclude false positives.  The curation process included two stages:
\begin{description}
\item[Stage 1:] Curation of \num{1000} software mentions using five
  subcategories: Software, Algorithm, Database, Web Platform, Hardware
  (and ``Other'').
  
  During this first stage of labeling, we learned that it is hard to
  distinguish between Software and Algorithms as these terms are often
  used interchangeably by researchers. For the next labeling stage we
  considered them both as ``Software'', whereas Database, Web Platforms
  and Hardware are considered as ``Non-Software''.

\item[Stage 2:] Curation of \num{9000} software mentions using two
  main categories: Software and Not-Software.
\end{description}

This document outlines the definitions used for the annotation
guidelines used by our domain experts for the manual labeling task,
including examples of mentions to include (and exclude) in the
dataset. The guidelines evolved through several iterative annotation
cycles, taking into account feedback and suggestions from the curators
where rules needed to be more explicit.

\subsection*{Definitions of subcategories}

Our SciBERT model was trained on the Softcite
Dataset \cite{Du2021}. We used their curation material as a baseline
for our definitions:
\begin{description}
\item[Software] includes all `obvious' software: a good indication for
  inclusion is if a tool can be downloaded and installed; however,
  ``Software as a Service''---platforms providing up-to-date
  cloud-based services for bioinformatic data analysis over a
  website---are also included in this category.  Programming
  languages, such as ``Java'' or ``R'', are annotated as ``Software''; this
  includes mention of a script written in a language (the Softcite
  coding scheme explains that ``programming languages are themselves
  software being used to create software'').  
\item[Algorithm] is defined as a program, a problem-solving process
  that is computerized, or a function. The programs and algorithms may
  have implementations in different languages. 
\item[Database] is a data collection or dataset, knowledgebase or
  repository. Note that in some cases, the database may be supported
  by a software framework (e.g. with the data made accessible via an
  interface); if the mention refers to the software component, it
  should be labeled as Software (often ``Software as a service'').  
\item[Web platform] includes web services such as ``Facebook'',
  ``Google'', or ``Amazon'', or other online platforms that have a web
  interface.  
\item[Hardware] is defined as physical instruments, devices,
  components or delivery systems, or other hardware (that may have
  software installed on it).
\end{description}

\subsection*{Curation---stage 1: How to annotate}

Annotate each Software mention with one of the 5 categories:
\begin{itemize}
\item Software,
\item Algorithm,
\item Database,
\item Web Platform,
\item Hardware,
\end{itemize}
or with ``Other'' or ``Unclear''.

\begin{enumerate}
\item For each Software mention 5 example sentences from papers are
  included. Use example sentences to understand the context and
  whether the mention is referred to as ``software'', ``version'',
  ``program'', etc.

\item If the context is not clear, look online for more information,
  e.g. the About section of an online tool, information provided in a
  GitHub entry, a paper abstract describing the ‘mention’.

\item Where possible, label the mention with one of the subcategories.
  Mentions that don’t fit any of the subcategories 1-5 should be
  labeled with ``Other'' (see examples below).

\item For mentions where more than one subcategory applies, use the
  following rules:
\begin{itemize}
\item Where the distinction between ``Software'' and ``Algorithm'' is
  difficult $\to$ select ``Software''.

\item Where the mention could be ``Software'' and another category
  $\to$ select ``Software''.

\item Similarly, where it could ``Algorithm'' and another category
  $\to$ select ``Algorithm''.  For example, ``VISTA is a comprehensive
  suite of programs and databases for comparative analysis of genomic
  sequences'' $\to$ label as ``Algorithm''.
   
\item Where the mention could be labeled with two other categories
  $\to$ select either. (Accuracy was deemed less important here as the
  final goal was to distinguish between Software+algorithm vs
  not-Software).
\end{itemize}
\item For ambiguous mentions and acronyms, where the example sentences
  appear to refer to multiple different things, select ``Unclear''.
\end{enumerate}

\subsection*{Curation---stage 2: How to annotate}

Annotate each Software mention with one of the two main categories:
\begin{itemize}
\item Software \& Algorithm,
\item Not Software (including databases, web platforms,
  hardware, or Other) Or with ``Unclear''
\end{itemize}

How to annotate: 
\begin{enumerate}
\item As before, use example sentences to understand the context and
  whether a mention is about ``software'', ``program'', ``algorithm''
  etc. If the context is not clear from the examples, look online for
  an online tool, a GitHub entry, a paper abstract, etc.    

\item For ambiguous mentions, where the sample sentences appear to
  refer to multiple different things and don’t fall clearly into one
  of two main categories, select ``Unclear''. 

\item True software\_mention that appear as partial terms should be
  labeled as Software. For example, the Software mention ``Random''
  can be labeled as Software if the example sentences clearly refer to
  the same software (``The superior accuracy of activity measures was
  confirmed using Random Forest and predictive modeling techniques'').
\end{enumerate}

\subsection*{Examples}

Examples of tools that should be labeled as ``Software \& Algorithm''
(Curation stage 2) are shown in Table~\ref{tab:examples_software}.  

\begin{table*}
  \centering
  \caption{Curation Guidelines - Examples of mentions: software \& algorithm. These examples were used for training our curation team.}
  \label{tab:examples_software}
  \begin{tabularx}{1.0\textwidth}{llXX}
    \toprule
    \multicolumn{1}{p{4em}}{Software mention ID}
    & \multicolumn{1}{p{4em}}{Software mention}
    & Link
    & Description at the link and notes\\
    \midrule
    SM2407
    & ABAQUS
    & \url{https://en.wikipedia.org/wiki/Abaqus}
    & \emph{``Abaqus FEA (formerly ABAQUS) is a software suite for finite
      element analysis and computer-aided engineering''}
      \\
    SM6358
    & gplots
    & \url{https://cran.r-project.org/web/packages/gplots/index.html}
    & \emph{``Various R programming tools for plotting data''}
      \\
    SM5899
    & Java
    & \url{}
    &
      \\
    SM1591
    & Perl
      \\
    SM8176
    & Keras
    & \url{https://keras.io/}
    & \emph{``Keras is an open-source software library that provides a Python
      interface for artificial neural networks''}
      \\
    SM1028
    & I-TASSER
    & \url{https://zhanggroup.org/I-TASSER/}
    & \emph{``server for protein structure and function prediction''}
      \\
    SM1500
    & GEPIA
    & \url{http://gepia.cancer-pku.cn/}
    & \emph{``interactive web server for analyzing the RNA sequencing
      expression data''}
      \\
    SM534
    & InterProScan
    & \url{https://www.ebi.ac.uk/interpro/about/interproscan/}
    & \emph{``InterProScan is the software package that allows sequences to
      be scanned against InterPro's member database signatures.''}
      \\
    SM14721
    & SurveyMonkey
    & \url{https://www.surveymonkey.co.uk/}
    & online questionnaire tool
      \\
    SM1693
    & ARB
    & \url{https://en.wikipedia.org/wiki/ARB_Project}
    & \emph{``The ARB Project is a free software package for phylogenetic
      analysis of rRNA''}
      \\
    SM6650
    & BiNGO
    & \url{https://www.psb.ugent.be/cbd/papers/BiNGO/Home.html}
    & \emph{``a Java-based tool to determine which Gene Ontology (GO)
      categories are statistically overrepresented in a set of genes
      or a subgraph of a biological network''}
      \\
    SM24091
    & ADMIXTURE
    & \url{https://bioinformaticshome.com/tools/descriptions/ADMIXTURE.html}
    & \emph{``Enhancements to the ADMIXTURE algorithm for individual ancestry
      estimation.''}
      \\
    SM34896
    & ABySS
    & \url{https://rcc.fsu.edu/software/abyss}
    & \emph{``ABySS is a Bioinformatics program designed to assemble genomes
      from small paired-end sequence reads''}
      \\
    SM8175
    & Adam
    & \url{https://towardsdatascience.com/adam-latest-trends-in-deep-learning-optimization-6be9a291375c}
    & \emph{``Adam is a replacement optimization algorithm for stochastic
      gradient descent for training deep learning models''}
      \\
    SM6044
    & Blast
    & \url{https://blast.ncbi.nlm.nih.gov/Blast.cgi}
    & \emph{``BLAST is an algorithm and program for comparing primary
      biological sequence information''}
      \\
    SM15716
    & bowtie
    & \url{https://bio.tools/bowtie2}
    & \emph{``Bowtie 2 is an ultrafast and memory-efficient tool for aligning
      sequencing reads to long reference sequences''}
      \\
    SM1735
    & TopHat
    & \url{https://ccb.jhu.edu/software/tophat/index.shtml}
    & \emph{``TopHat-Fusion algorithm''}
      \\
    SM4671
    & MySQL
    & \url{https://en.wikipedia.org/wiki/MySQL}
    & open-source relational database management system
      \\
    SM3262
    & ClueGO
    & \url{https://apps.cytoscape.org/apps/cluego}
    & Software plugin
      \\
    SM4915
    & R script
    &
    & Software script
    \\
    SM5658
    & CUDA
    & \url{https://docs.nvidia.com/cuda/}
    & API
      \\
    SM517
    & DNASTAR
    & \url{https://www.dnastar.com/}
    & software that has the same name as the company that develops it
      \\
    SM1004
    & FASTA 
    & \url{https://en.wikipedia.org/wiki/FASTA_format}
    & software that has the same name as the output format (if mention is clearly the software)
      \\
\bottomrule
  \end{tabularx}
\end{table*}

Examples of tools that should be labeled as ``Not-Soft\-ware'' (Curation
stage 2) are shown in Table~\ref{tab:examples_not_software}. The
different types (subcategories) used for labeling in stage 1 are
indicated in the last column.

\begin{table*}
  \centering
  \caption{Curation Guidelines - Examples of mentions: not-software. These examples were used for training our curation team.}
  \label{tab:examples_not_software}
  \begin{tabularx}{1.0\linewidth}{llXXl}
    \toprule
    \multicolumn{1}{p{4em}}{Software mention ID}
    & \multicolumn{1}{p{4em}}{Software mention}
    & Link
    & Description at the link and notes
    & Subcategory
    \\
    \midrule
    SM37528
    & WorldClim
    & \url{https://www.worldclim.org/data/v1.4/worldclim14.html}
    & \emph{``Maps, graphs, tables, and data of the global climate''}
    & Database
    \\
    SM15080
    & ArrayExpress
    & \url{https://www.ebi.ac.uk/arrayexpress/}
    & Data archive
    & Database
    \\
    SM8406
    & cgMLST
    & \url{https://www.cgmlst.org/ncs}
    & \emph{``Nomenclature Server''}
    & Database
    \\
    SM15100
    & ClinVar
    & \url{https://www.ncbi.nlm.nih.gov/clinvar/}
    & \emph{``ClinVar aggregates information about genomic variation''}
    & Database
    \\
    SM2352
    & GitHub
    & \url{https://github.com/}
    & Software repository
    & Database
    \\
    SM3888
    & Facebook
    & 
    & \emph{``online social media and social networking service''}
    & Web platform
    \\
    SM5608
    & Google Earth
    & 
    & \emph{``Google Earth is a geobrowser''}
    & Web platform
    \\
    SM5472
    & Google Play
    & \url{https://play.google.com/about/howplayworks/}
    & \emph{``Google Play is an online store where people go to find
            their favorite apps, games''}
    & Web platform
    \\
    SM5379
    & ActiGraph
    & \url{https://actigraphcorp.com/}
    & \emph{``ActiGraph’s wearable ac\-ce\-le\-ro\-meter-based biosensors''}
    & Hardware
    \\
    SM3147
    & Leica
    & \url{https://leica-camera.com/en-GB}
    & Company name and product name
    & Hardware
    \\
    SM5200	
    & Kinect
    & \url{https://en.wikipedia.org/wiki/Kinect}
    & \emph{``Kinect is a line of motion sensing input devices produced by
      Microsoft''} 
    & Hardware
    \\
\bottomrule
  \end{tabularx}
\end{table*}

Examples of mentions that don’t fit into the five subcategories are
liested in Table~\ref{tab:examples_other}. They should be labeled as
``Not-Software'' in Curation stage 2 (or as ``Other'' in stage 1). 

\begin{table*}
  \centering
  \caption{Curation Guidelines - Examples of mentions: ``other'' mentions.  These examples were used for training our curation team.}
  \label{tab:examples_other}
  \begin{tabularx}{1.0\linewidth}{llXXp{0.15\textwidth}}
    \toprule
    \multicolumn{1}{p{4em}}{Software mention ID}
    & \multicolumn{1}{p{4em}}{Software mention}
    & Link
    & Description at the link
    & Notes
    \\
    \midrule
    SM5111
    & ANOSIM
    & \url{https://www.nhm.uio.no/english/research/infrastructure/past/help/anosim.html}
    & ANOSIM (ANalysis Of Similarities) is a non-parametric test of significant difference
    & Statistical test
    \\
    SM1779
    & LOESS
    & \url{https://www.statsdirect.com/help/nonparametric_methods/loess.htm}
    & method for fitting a smooth curve between two variables
    & Statistical test
    \\
    SM5147
    & ENCODE
    & \url{https://www.encodeproject.org/}
    & public research project which aims to identify functional elements in the human genome
    & Consortium/pro\-ject
    \\
    SM1166
    & R Core
    & \url{https://www.r-project.org/contributors.html}
    & a core group, the R Core Team, with write access to the R source
    & Group of people
    \\
    SM1901	
    & \multicolumn{1}{p{5em}}{\RaggedRight R Foundation for Statistical Computing}
    & \url{https://www.r-project.org/foundation/}
    & The R Foundation is a not for profit organization
    & Group of people
    \\
    SM3954
    & RE-AIM
    & \url{https://re-aim.org/}
    & RE-AIM is a framework to guide the planning and evaluation of programs according to the 5 key RE-AIM outcomes: Reach, Effectiveness, Adoption, Implementation, and Maintenance
    & Project
    \\
    SM5429
    & iOS
    & 
    & 
    & Operating system
    \\
    SM6157
    & Affymetrix
    & \url{https://www.thermofisher.com/uk/en/home/life-science/microarray-analysis/affymetrix.html}
    & \emph{``Affymetrix microarray solutions include necessary components for a microarray experiment, from arrays and reagents to instruments and software''}
    & System with multiple components, including software
    \\
    SM16352
    & nCounter 
    & \url{https://nanostring.com/products/ncounter-analysis-system/ncounter-pro/}
    & 
    & System with multiple components, including software
    \\
    SM3461
    & Arduino
    & \url{https://www.arduino.cc/}
    & 
    & System with multiple components, including software
    \\
    SM1127
    & MiSeq
    & \url{https://emea.illumina.com/systems/sequencing-platforms/miseq.html}
    & \emph{``applications such as targeted resequencing, metagenomics, small genome sequencing, targeted gene expression profiling''}
    & System with multiple components, including software
    \\
    SM8468
    & Apache
    & \url{https://www.apache.org/licenses/LICENSE-2.0}
    & 
    & License
    \\
    SM338
    & QUADAS
    & \url{https://www.bristol.ac.uk/population-health-sciences/projects/quadas/quadas-2/}
    & A quality assessment tool for diagnostic accuracy studies
    & Tools that are checklists
    \\
    SM1908
    & RNAseq
    &     
    & sequencing technique
    & Method
    \\
    SM5014
    & SOLiD
    & 
    & SOLiD next-generation sequencing technology
    & Method
    \\
\bottomrule    
  \end{tabularx}
\end{table*}

\section{Datasets Description}
\label{sec:results}

In this section, we provide details about the datasets we make
available.

There are six directories: \path{raw}, \path{disambiguated},
\path{linked}, \path{evaluation} and \path{intermediate}.  Files with
the extension \path{tsv} are tab separated, files with the extension
\path{csv} are comma separated, files with the suffix \path{pkl} are
Python serialized objects~\cite{van1995python}, files with the suffix
\path{gz} are gzipped.  Note that tab separated files \emph{may}
contain embedded quotes, which do not have special meaning, while in
comma separated files they do according to the usual conventions.

\subsection{Raw files}
\label{sec:raw}

The \path{raw} directory contains extracted mentions before
disambiguation and linking.  It has the following three files:
\path{comm_raw.tsv.gz} (PMC OA commercial subset),
\path{non_comm_raw.tsv.gz} (PMC OA non-commercial subset), and
\path{publishers_collection_raw.tsv.gz} (CZ Publishers' collection).
Each file also contains a \textbf{curation\_label} field to denote the curation label our curation team gave to each software entry. 

The files \path{comm_raw.tsv.gz} (commercial subset) and
\path{non_comm_raw.tsv.gz} (non-commercial subset) are tab
separated, gzipped, and have the following fields:
\begin{description}
\item[license] either \path{comm} or \path{non_comm},
\item[location] the location of the file from which the mentions are
  extracted, for example, \path{comm/Micropl/PMC8475362.nxml},
\item[pmcid] PMC id of the paper (with the prefix ``PMC'' stripped),
\item[pmid] PubMed id of the paper,
\item[doi] DOI of the paper, 
\item[pubdate] publication year according to the metadata in the paper source,
\item[source] part of the paper from which the mention was extracted; this is either a section in the main text of the paper (like "introduction" or "materials and methods") or another part of the paper: \path{paper_title},
  \path{paper_abstract}, \path{tab_caption}, \path{fig_caption}.
\item[number] sequence number of the object, from which the mention
  was extracted:
  \begin{itemize}
  \item for body text, paragraph number;
  \item for figure and table captions, the number of the figure or
    table;
  \item otherwise, zero,
  \end{itemize}
\item[text] the sentence, from which the software mention was
  exracted,
\item[software] the extracted software mention,
\item[version] the extracted software version,
\item[ID] software mention ID.
\item[curation\_label] curation result for the software mention: 
\begin{itemize}
  \item \textbf{software} if the mention was labeled as software by the curation team
  \item \textbf{not\_software} if labeled as not-software by the curation team
  \item \textbf{unclear} if a call could not be made based on available information
  \item \textbf{not\_curated} if mention has not been curated.
\end{itemize}
\end{description}

Results of extraction from the CZ Publishers' collection are in the file
\path{publishers_collection.tsv.tgz}.  It is a tab separated gzipped
file with the following fields:
\begin{description}
\item[doi] paper DOI,
\item[pubdate] publication year, according to the metadata in the paper source,
  \item[source] part of the paper from which the mention was extracted; this is either a section in the main text of the paper (like "introduction" or "materials and methods") or another part of the paper: \path{paper_title},
    \path{paper_abstract}, \path{tab_caption}, \path{fig_caption}.
  \item[number] sequence number of the object, from which the mention
    was extracted:
    \begin{itemize}
    \item for body text, paragraph number;
    \item for figure and table captions, the number of the figure or
      table;
    \item otherwise, zero,
    \end{itemize}
  \item[text] the sentence, from which the software mention was
    exracted,
  \item[software] the extracted software mention.
  \item[ID] software mention ID.
  \item[curation\_label] curation result for the software mention: 
    \begin{itemize}
      \item \textbf{software} if the mention was labeled as software by the curation team
      \item \textbf{not\_software} if labeled as not-software by the curation team
      \item \textbf{unclear} if a call could not be made based on available information
      \item \textbf{not\_curated} if mention has not been curated.
    \end{itemize}
\end{description}

\subsection{Disambiguation Results}
\label{sec:disambiguated}

The \path{linked} directory contains the directory
\path{synonyms_files} and the files \path{comm_disambiguated.tsv.gz}.

The \path{synonyms_files} directory contains synoynms dictionaries
used in the disambiguation process, stored in the pickle format. The
following files are available:
\begin{description}
\item[pypi\_synonyms.pkl] Python dictionary mapping from a pypi
  package and an array of synonyms generated through the Keywords
  Synonym Generation process \ref{sec:keywords-synonym-generation}
\item[cran\_synonyms.pkl] Python dictionary mapping from a CRAN
  package and an array of synonyms generated through the Keywords
  Synonym Generation process \ref{sec:keywords-synonym-generation}
\item[bioconductor\_synonyms.pkl] Python dictionary mapping from a
  Bioconductor package and an array of synonyms generated through the
  Keywords Synonym Generation process
  \ref{sec:keywords-synonym-generation}
\item[scicrunch\_synonyms.pkl] Python dictionary mapping from a
  mention found in SciCrunch and its synonyms retrieved through the
  SciCrunch API \ref{sec:scicrunch-synonyms-retrieval}
\item[extra\_scicrunch\_synonyms.pkl] Python dictionary mapping from a
  mention found in SciCrunch and its synonyms retrieved by parsing the
  corresponding URL in SciCrunch
  \ref{sec:scicrunch-synonyms-retrieval}
\item[string\_similarity\_synonyms.pkl] Python dictionary mapping from
  a mention found in the \path{comm_IDs.tsv.gz} corpus and its
  synonyms retrieved through the Jaro Winkler algorithm, together with
  the corresponding confidences
  \ref{sec:string-similarity-algos}. Only pairs of synonyms with a
  similarity confidence of $\geq 0.9$ are kept. The file has the
  following format:
  \begin{itemize}
  \item software\_mention - mention synonyms are computed for
  \item ([synonyms], [synonyms\_confidences]) - tuple containing two arrays:
    \begin{itemize}
    \item synonyms: list of synonyms for software mention
    \item synonyms\_confidences: Jaro Winkler similarity scores
      between synonyms and software mention, as given by the textdistance python package~\cite{textdistance}
    \end{itemize}
  \end{itemize}
\item[synonyms.tsv.gz] tab separated, gzipped comma file used as input
  for the clustering algorithm, after post-pro\-ces\-sing. The file is
  built by combining information from files \path{pypi_synonyms.pkl},
  \path{cran_synonyms.pkl}, \path{bioconductor_synonyms.pkl},
  \path{scicrunch_synonyms.pkl}, \path{extra_scicrunch_synonyms.pkl},
  and \path{string_similarity_synonyms.pkl}, and performing additional
  clean-up. The file has the following format:
  \begin{itemize}
  \item \textbf{ID} ID for software\_mention,
  \item \textbf{synonym\_ID} ID for synonym,
  \item \textbf{software\_mention} software\_mention to compute synonyms for
  \item \textbf{synoynm} a synonym for software\_mention
  \item \textbf{synonym\_conf} confidence for this synonym pair (as described in \ref{eq:Mij})
  \item \textbf{synonym\_source} source for this synonym pair:
    \begin{itemize}
    \item \textbf{SciCrunch} if synonym pair retrieved from \path{scicrunch_synonyms.pkl}
    \item \textbf{scicrunch\_page\_query} if synonym pair retrieved from \path{extra_scicrunch_synonyms.pkl}
    \item \textbf{Bioconductor} if synonym pair retrieved from \path{bioconductor_synonyms.pkl}
    \item \textbf{CRAN} if synonym pair retrieved from \path{cran_synonyms.pkl}
    \item \textbf{PyPI} if synonym pair retrieved from \path{pypi_synonyms.pkl}
    \item \textbf{string\_similarity} if synonym pair retrieved from \path{string_similarity_synonyms.pkl}
    \end{itemize}
    \end{itemize}
  \end{description}

  \subsubsection{Disambiguated File}

  The \path{comm_disambiguated.tsv.gz} file is a tab separated,
  gzipped file corresponding to the disambiguation results for
  \path{comm.tsv.gz}. The file was obtained by running the
  DBSCAN-based clustering algorithm on \path{synonyms.csv} as
  described in \ref{alg:clustering}. The file has all the fields in
  \path{comm.tsv.gz} with the additional fields:
  \begin{description}
  \item [mapped\_to\_software] software entity (or cluster) the software mention described in this entry is predicted to be part of; For example, for the software\_mention 'R package limma', the mapped\_to\_software might be `limma`.
  \item [mapped\_to\_software\_ID] ID of mapped\_to\_software
  \end{description}
  
\subsection{Linking Results}

The \path{linked} directory contains the following directories: \path{normalized} and \path{raw} and the file \path{metadata.tsv.gz}. 

\subsubsection{Raw Metadata Files}
The \path{raw} directory contains raw metadata files obtained by querying the PyPI, CRAN, Bioconductor, SciCrunch and GitHub APIs on mentions extracted by the NER algorithm from the \path{comm.tsv.gz} corpus (PMC-OA commercial subset). The directory contains the following comma separated files:
\begin{description}
\item[bioconductor\_raw\_df.csv] raw metadata file obtained by querying Bioconductor on mentions extracted from the \path{comm.tsv.gz}. Has fields:
    \begin{itemize}
        \item Bioconductor Package
        \item Bioconductor Link
        \item Maintainer
        \item Title
    \end{itemize}
\item[cran\_raw\_df.csv] raw metadata file obtained by querying CRAN on mentions extracted from the \path{comm.tsv.gz}. Has fields:,
    \begin{itemize}
        \item CRAN Package
        \item CRAN Link
        \item Title
    \end{itemize}
\item[github\_raw\_df.csv] raw metadata file obtained by querying GitHub on mentions extracted from the \path{comm.tsv.gz}. Has fields:,
    \begin{itemize}
        \item software\_mention
        \item best\_github\_match
        \item description
        \item github\_url
        \item license
        \item exact\_match
    \end{itemize}
\item[pypi\_raw\_df.csv] raw metadata file obtained by querying PyPI on mentions extracted from the \path{comm.tsv.gz}. Has fields:,
    \begin{itemize}
        \item pypi package
        \item pypi\_url
    \end{itemize}
\item[scicrunch\_raw\_df.csv] raw metadata file obtained by querying SciCrunch on mentions extracted from the \path{comm.tsv.gz}. Has fields:
    \begin{itemize}
        \item software\_name
        \item scicrunch\_synonyms
        \item Resource Name
        \item Resource Name Link
        \item Description
        \item Keywords
        \item Resource ID
        \item Resource ID Link
        \item Proper Citation
        \item Parent Organization
        \item Parent Organization Link
        \item Related Condition
        \item Funding Agency
        \item Relation
        \item Reference
        \item Website Status
        \item Alternate IDs
        \item Alternate URLs
        \item Old URLs
        \item Reference Link
    \end{itemize}
\end{description}

\subsubsection{Normalized Metadata Files}
The \path{normalized} directory contains normalized versions of the raw metadata files. Files are normalized to a common schema as described in \ref{tab:schema_normalization}. The directory contains the following comma separated files:
\begin{description}
\item[bioconductor\_df.csv] normalized metadata file obtained by querying Bioconductor on mentions extracted from the \path{comm.tsv.gz},
\item[cran\_df.csv] normalized metadata file obtained by querying CRAN on mentions extracted from the \path{comm.tsv.gz},
\item[github\_df.csv] normalized metadata file obtained by querying GitHub on mentions extracted from the \path{comm.tsv.gz}, 
\item[pypi\_df.csv] normalized metadata file obtained by querying PyPI on mentions extracted from the \path{comm.tsv.gz},
\item[scicrunch\_df.csv] normalized metadata file obtained by querying SciCrunch on mentions extracted from the \path{comm.tsv.gz}
\end{description}

\subsubsection{Master Metadata File}
The \path{metadata.tsv.gz} file is a concatenation of all the metadata files in the \path{normalized} directory. Each file in the \path{normalized} directory, as well as the \path{metadata.tsv.gz} file has the following fields:

\begin{description}
\item[ID] a unique identificatior for each software,
\item[software\_mention] the canonical string for the given software
  mention,
\item[mapped\_to] list of values to which the software was mapped,
\item[source] mapping source (PyPI, CRAN, SciCrunch, GitHub, Bioconductor),
\item[platform] list of platforms for the given software,
\item[package\_url] URL for the given package,
\item[description] list of descriptions associated with software in
  the databases,
\item[homepage\_url] list of homepages for the software,
\item[other\_urls] list of other URLs for the given software mined
  from the database,
\item[license] list of licenses under which the software
  is released,
\item[github\_repo]  list of GitHub repositories for the software,
\item[github\_repo\_licenses] list of licenses listed on the GitHub
  repositories,
\item[exact\_match] \path{True} if an exact string match was found
  for the given software mention, \path{False} if a fuzzy match was
  used instead,
\item[RRID]  RRID for the software retrieved from
  SciCrunch~\cite{SciCrunch}, 
\item[reference] journal articles linked to the software, identified
  by DOI, PMID or RRID,
\item[scicrunch\_synonyms] synonyms for software according to
  SciCrunch~\cite{SciCrunch}.
\end{description}

\subsection{Evaluation Files}
\label{sec:evaluation_data}

The directory \path{evaluation} contains files used for curation and evaluation.

\subsubsection{Curated Software Mentions}

The file \path{curation_top1k_mentions_multi_labels.csv.gz} contains
the results of curation of top \num{10000}~mentions over five
categories.  It is a comma separated file with the following fields:
\begin{description}
\item [ID] unique identifier for software mention,
  
\item [software_mention] name of software mention,
  
\item [text] five example sentences from articles where the software
  mention appears,
  
\item [multi\_label] manually curated label selected for the software
  mention. Possible labels: \textbf{software, algorithm, database, hardware,
  web platform, other, unclear},
  
\item [label] manually curated label selected for the software
  mention. Possible labels: software\&algorithm, 
  not_software, unclear,
  
\item [Curation\_comments] curator’s explanation why a specific label
  was chosen, including links to online tools, spelled-out acronyms,
  information found through online searches and/or in example
  sentences.
  
\end{description}

The file \path{curation_top10k_mentions_binary_labels.csv.gz} contains 
the results of curation of top \num{1000}~mentions over two
categories.  It is a comma separated file with the following fields:
\begin{description}
\item [ID] unique identifier for software mention,
  
\item [software\_mention] name of software mention,
  
\item [text] five example sentences from articles where the software
  mention appears,
  
\item [pmid] PubMed identifiers for the articles from which the five
  example sentences (in column “text”) were extracted,
  
\item [pmcid] PubMed Central identifiers for the articles,
  
\item [doi] digital object identifier for the articles,
  
\item [label] manually curated label selected for the software
  mention. Possible labels: software\&algorithm, 
  not_software, unclear,
  
\item [curation\_notes] curators' explanation why a specific label
  was chosen, including links to online tools, spelled-out acronyms,
  information found through the online searches and/or in example
  sentences.
  
\end{description}

\subsubsection{Linking \& Disambiguation Evaluation Files}
The file \path{evaluation_linking.csv.gz} presents the manual
evaluation of linking. Curators were tasked with assessing if the link generated
for a particular \textbf{software\_\allowbreak mention} was correct. The file is a gzipped comma-separated 
file with the same fields as \path{comm_curated.tsv.gz} and \path{metadata.tsv.gz} above, 
and an additional field:

\begin{description}
\item [software\_mention] manually assigned label by the curators:
\begin{itemize}
\item \textbf{correct} if the \textbf{package\_url} generated for
  \textbf{software\_mention} is correct;
    \item \textbf{inccorrect} if the \textbf{package\_url} generated for \textbf{software\_mention} is incorrect;
    \item \textbf{unclear} if there isn't enough information to assess
      whether or not the \textbf{package\_url} generated for
      \textbf{software\_mention} is correct;
    \end{itemize}
\end{description}

The file \path{evaluation_disambiguation.csv.gz} presents the manual
evaluation of disambiguation. Curators were tasked with asssessing if a pair of string software mentions are synonyms. The file is a gzipped comma-separated file with the following fields:
\begin{description}
\item [link\_label] name of software mention,
\item [synonym] synonym generated by the disambiguation algorithm,
\item [text] five example sentences from articles where the synonym
  mention appears,
  
\item [pmid] PubMed identifiers for the articles from which the five
  example sentences (in column “text”) were extracted,
  
\item [pmcid] PubMed Central identifiers for the articles,
  
\item [doi] digital object identifier for the articles,
  
\item [synonym_label] manually curated label selected for the \textbf{software\_mention, synonym pair}. Possible labels: 'Exact', 'Narrow', 'Not software', 'Unclear', 'Not synonym'
  
\item [curation\_notes] curators' explanation why a specific label
  was chosen
\end{description}

\subsection{Intermediate Files}
\label{sec:intermediate_files}

The directory \path{intermediate_files} contains the following files: \path{mention2ID.pkl} and \path{freq_dict.pkl}

The file \path{mention2ID.pkl} is a \textbf{mention} to \textbf{ID} mapping connecting all the plain text software mentions extracted by the NER algorithm from the \path{comm_raw.tsv.gz},
\path{non_comm_raw.tsv.gz} and
\path{publishers_collection_raw.tsv.gz} to a unique ID. The file is in the pickle format and the data is stored as a Python dictionary. The keys are plain-text software mentions across all three datasets, and the values are unique IDs across all three corpora. 
\begin{description}
\item[mention] plain-text software mention
\item[ID] unique ID for the plain-text software mention
\end{description}

The file \path{freq_dict.pkl} is a \textbf{mention} to \textbf{frequency} mapping connecting all plain text software mentions extracted by the NER algorithm from the \path{comm_raw.tsv.gz} to the their frequency on the \path{comm_raw.tsv.gz} dataset. We define frequency as the total number of unique papers a mention appears in the PMC-OA \textsl{comm} dataset. The file is in the pickle format and the data is stored as a Python dictionary. The keys are plain-text software mentions in  \path{comm_raw.tsv.gz}, and the values are the corresponding frequencies. 
\begin{description}
\item[mention] plain-text software mention
\item[frequency] total number of unique papers in the \path{comm_raw.tsv.gz} dataset the mention appears in
\end{description}

\section{Additional Examples of String Variability}

\label{sec:string_variability}

Additional examples of string variability can be found on
Figures~\ref{fig:matlab_variability}
and~\ref{fig:graphpad_variability}.  

\clearpage
\begin{figure}
  \hrulefill\\
    matrix laboratory \\
    MATLAB \\
    MATLAB, Signal Processing Toolbox and Statistics Toolbox \\
    MATLAb \\
    MATLab \\
    MATlab \\
    MaTLab \\
    MatLAB \\
    MatLab \\
    Matlab \\
    MATLAB, Statistics and Machine Learning Toolbox \\
    mAtLab) \\
    Matlab) \\
    matlab \\
    matlab - the language of technical computing \\
    matlab) \\
    matlab, signal processing toolbox and statistics toolbox \\
    matlab, statistics and machine learning toolbox \\
    MATLAB \\
    matrix laboratory \\
    matlab \\
    MATLAB -The Language of Technical Computing \\
    MATLAB BGL \\
    MATLAB IDE \\
    MATLAB HMM \\
    MATLAB GUI \\
    MATLAB GPU \\
    MATLAB FDA \\
    MATLAB EFD \\
    MATLAB DIC \\
    MATLAB DGN \\
    MATLAB CVX \\
    MATLAB CPU \\
    MATLAB App \\
    MATLAB® SVM \\
    MATLAB ANN \par
    \hrulefill\par
  \caption[Examples of string variability for the software
  `MATLAB']{Examples of string variability for the software `MATLAB'
    (\url{http://www.mathworks.com/products/matlab/}), as extracted by
    the NER model from the PMC-OA corpus}
  \label{fig:matlab_variability}
\end{figure}

\begin{figure}
  \hrulefill\\
    graphpad, prism \\
    GraPhPad \\
    GraphPAD \\
    GraphPAd \\
    GraphPad \\
    GraphPad) \\
    GRAPHPAD \\
    graphpad \\
    graphpad) \\
    GraphPad, Prism \\
    GraphPad, Prism \\
    GraphPad- Prism \\
    GraphPad-InSTat \\
    GraphPad-InStat \\
    GraphPad-Instat \\
    GraphPad-Prism4 \\
    GraphPad-Prism8 \\
    GraphPad-Prism7 \\
    GraphPad-Prism© \\
    GraphPad-prism5 \\
    GraphPad “Prism \\
    GraphPad-prism7 \\
    GraphPad-Prism5 \\
    GraphPad “PRISM \\
    GraphPad prism8 \\
    GraphPad ©Prism \\
    GraphPad prism® \\
    GraphPad prisma \\
    GraphPad prism9 \\
    GraphPad.PRISM® \\
    GraphPad prism7 \\
    GraphPad prism6 \\
    GraphPad prism5 \\
    GraphPad prism4 \\
    GraphPad prism3 \\
    GraphPad primer \\
    GraphPad instat \\
    GraphPad) \par
    \hrulefill\par
    \caption[Examples of string variability for the software `GraphPad']{Examples of string variability for the software `GraphPad'
    (\url{http://graphpad.com/}), as extracted by the NER model from
    the PMC-OA corpus}
  \label{fig:graphpad_variability}
\end{figure}

\end{document}